\documentclass[twocolumn,showpacs,amsmath,amssymb]{revtex4}
\usepackage{graphicx}
\usepackage{enumerate}
\usepackage{dcolumn}
\usepackage{bm}

\begin{document}

\title{Cluster-based architecture
for fault-tolerant quantum computation}

\author{Keisuke Fujii}
\author{Katsuji Yamamoto}
\affiliation{
Department of Nuclear Engineering, Kyoto University, Kyoto 606-8501, Japan}

\date{\today}

\begin{abstract}
We present a detailed description of an architecture
for fault-tolerant quantum computation,
which is based on the cluster model of encoded qubits.
In this cluster-based architecture,
concatenated computation is implemented in a quite different way
from the usual circuit-based architecture
where physical gates are recursively replaced by logical gates
with error-correction gadgets.
Instead, some relevant cluster states, say fundamental clusters,
are recursively constructed through verification and postselection
in advance for the higher-level one-way computation,
which namely provides error-precorrection of gate operations.
A suitable code such as the Steane seven-qubit code
is adopted for transversal operations.
This concatenated construction of verified fundamental clusters
has a simple transversal structure of logical errors,
and achieves a high noise threshold $ \sim 3 \% $ for computation
by using appropriate verification procedures.
Since the postselection is localized within each fundamental cluster
with the help of deterministic bare controlled-$Z$ gates
without verification, divergence of resources is restrained,
which reconciles postselection with scalability.
\end{abstract}

\pacs{03.67.Lx, 03.67.Pp, 03.67.-a}

\maketitle

\section{Introduction}
\label{sec:introduction}

In order to implement reliable computation in physical systems,
either classical or quantum, the problem of noise should be overcome.
Particularly, fault-tolerant schemes have been developed
based on error correction in quantum computation
\cite{Shor95,CaldShor96,Shor96,Stea96,DiViShor96,Stea97,Stea98,Gottesman97}.
In the usual quantum error correction (QEC), error syndromes are detected
on encoded qubits, and the errors are corrected according to them.
The noise thresholds for fault-tolerant computation are calculated
to be about $10^{-6} - 10^{-3}$
depending on the QEC protocols and noise models
\cite{Stea97,Stea98,Gottesman97,Stea99,Stea03,Kitaev97,Preskill98,Knill98,
Gottesman98,AB-O99}.
A main motivation for QEC comes from the fact that
in the circuit model the original qubits should be used
throughout computation even if errors occur on them.

On the other hand, more robust computation may be performed
in measurement-based quantum computers
\cite{GC99,ZLC00,OWC,Niel03,Tame07,Knill05a,Knill05b}.
Teleportation from old qubits to fresh ones
is made by measurements to implement gate operations,
and the original qubits are not retained.
An interesting fault-tolerant scheme
with error-correcting teleportation is proposed
based on encoded Bell pair preparation and Bell measurement,
which achieves high noise thresholds $ \sim 3 \% $
\cite{Knill05a,Knill05b}.
The cluster model or one-way computer \cite{OWC}
should also be considered for fault-tolerant computation.
A highly entangled state, called a cluster state, is prepared,
and gate operations are implemented
by measuring the qubits in the cluster
with feedforward for the postselection of measurement bases.
This gate operation in the cluster model
may be viewed as the one-bit teleportation \cite{ZLC00}.
A promising scheme for linear optical quantum computation is proposed,
where deterministic gates are implemented
by means of the cluster model \cite{Niel04}.
Fault-tolerant computation is built up for this optical scheme
by using a clusterized version of the syndrome extraction for QEC
\cite{Stea97}.
The noise thresholds are estimated to be about $ 10^{-3} $ for photon loss
and $ 10^{-4} $ for depolarization \cite{Niel06}.
The threshold result is also argued
by simulating the QEC circuits with clusters
\cite{Rausen03,ND05,AL06}.
Furthermore, topological fault-tolerance in cluster-state computation
is investigated in a two-dimensional nearest-neighbor architecture,
where a high noise threshold $ \sim 0.75 \% $ is obtained
in spite of its strong physical constraint \cite{Raussendorf07}.
Some direct approaches are, on the other hand, considered
for the fault-tolerant one-way computation
\cite{FY07,Silva07,JF09},
though there seems to be a problem for scalability.

In this paper we present a systematic and comprehensive description
of an architecture for fault-tolerant quantum computation,
namely the cluster-based architecture,
which has been proposed recently to reconcile postselection with scalability
by virtue of one-way computation \cite{FY09}.
Specifically, the fault-tolerant computation
is implemented by concatenated construction and verification
of logical cluster states
via one-way computation with postselection.
A number of cluster states are constructed
in parallel with error detection,
and the unsuccessful ones are discarded,
selecting the clean cluster states.
The error-correcting teleportation (or its cluster version)
\cite{Knill05a,Knill05b,Silva07,JF09}
requires a high-fidelity preparation of Bell state.
It is also considered that improved ancilla preparation
increases the noise threshold \cite{Reichardt04,Eastin07}.
In the present cluster-based architecture \cite{FY09},
even gate operations as logical cluster states
are prepared and verified by postselecting the lower-level computation
to reduce the errors efficiently
(see also Ref. \cite{FY07} for an early idea).
This is quite distinct from the usual circuit-based QEC architectures,
including the error-correcting teleportation,
where the errors are corrected after noisy gate operations.

While high-fidelity state preparation is achieved by postselection,
huge resources are generally required
due to the exponentially diminishing net success probability
according to the computation size.
This is a serious obstacle for scalability
in the postselecting schemes
\cite{Knill05a,Knill05b,FY07}.
Here, we succeed in overcoming this problem in postselection
by presenting a systematic method of concatenation
to construct logical cluster states through verification,
where the unique feature of the cluster-model computation
is fully utilized.
As described in detail later, the necessary postselections
are minimized and localized
by dividing a whole cluster state into some fundamental clusters
with the help of controlled-$ Z $ (C$Z$) gates without verification,
say bare C$Z$ gates.
This enables the off-line gate operations prior to the computation
as the verified logical cluster states,
and provides a scalable concatenation with postselection
in the cluster-model computation.
The concatenated construction of verified clusters
is implemented with transversal (bitwise) operations
by adopting a suitable code such as the Steane seven-qubit code,
which belongs to a class of stabilizer codes of Calderbank-Shor-Steane (CSS)
\cite{CaldShor96,Stea96,Gottesman98}.
The logical measurements of Pauli operators
as well as the Clifford gates, $ H $, $ S $ and C$Z$,
are implemented transversally on such a quantum code.
The non-Clifford $ \pi / 8 $ gate is even operated
for universal computation by preparing a specific qubit
and making a transversal measurement \cite{FY07,Silva07}.
By exploiting this good transversal property,
the cluster-based architecture has a simple structure of logical errors
in concatenation to estimate readily the noise threshold.
A high noise threshold $ \sim 3 \% $ can be achieved
by using appropriate verification procedures with postselection.
Furthermore, the resources usage is moderate,
being comparable to or even less than
those of the circuit-based QEC architectures.

The rest of the paper is organized as follows.
In Sec. \ref{sec:back ground} we briefly review
the usual fault-tolerant quantum computation with circuit-based QEC.
In Sec. \ref{sec:simple model}
we introduce the main concept of cluster-based architecture
by considering a simple model preliminarily.
In Sec. \ref{sec:concatenated construction}
we present a detailed description of an efficient architecture
for the concatenated construction of verified logical clusters.
The fundamental clusters and verification protocols
are suitably adopted there, namely the hexa-cluster, code states,
single and double verifications.
Then, performance of the architecture is analyzed
in Secs. \ref{sec:threshold}, \ref{sec:resources}
and \ref{sec:miscellaneous},
with respect to the noise threshold and resources usage.
Sec. \ref{sec:conclusion} is devoted to summary and conclusion.
In the Appendix \ref{sec:code state}
we explain how to produce the cluster diagrams
to construct the fundamental clusters with single and double verifications.

\section{Circuit-based fault-tolerant architecture}
\label{sec:back ground}

We first review the usual fault-tolerant architecture
based on the circuit-model computation with QEC.
In comparison, this will be helpful to understand the distinct feature
of the cluster-based fault-tolerant architecture,
which will be investigated in the succeeding sections.

It is well known that by using QEC codes
we can protect quantum information from errors
which are caused by interaction with environment.
Specifically, by adopting the stabilizer codes
we can perform syndrome detection for recovery operation
simply by measuring the stabilizer operators.
Several QEC gadgets have been proposed
to implement the stabilizer measurement in a fault-tolerant way
\cite{DiViShor96,Stea97,Knill05a}.
A QEC gadget was first proposed by DiVincenzo and Shor,
where cat states are used as ancillae for the syndrome measurement
\cite{DiViShor96}.
Subsequently, a relatively simple type of QEC gadget was proposed
by Steane \cite{Stea97}, where encoded ancilla states are used
to extract the syndrome with transversal operations.
Especially, in the case of CSS code
the logical code states can be used as ancilla states.
For example, the following circuit
executes the $Z$ and $X$ error syndrome extractions
by using the ancilla $ | 0_L \rangle $ states,
\begin{equation}
\scalebox{.4}{\includegraphics*[0cm,0.5cm][16cm,6cm]{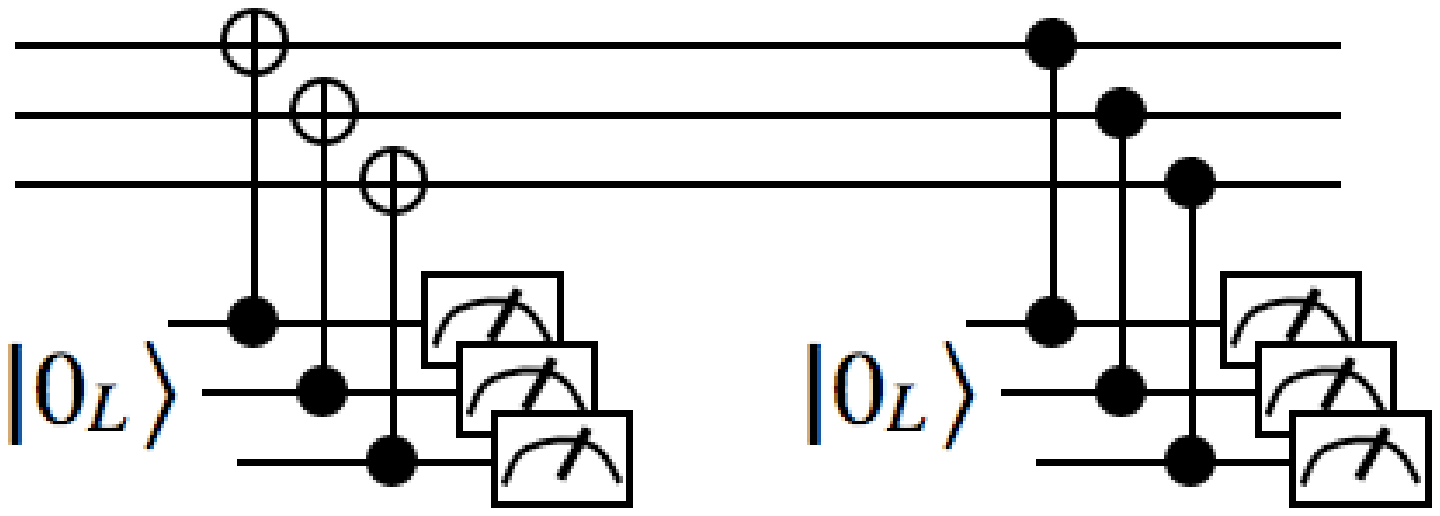}}
\label{eqn:QEC1}
\end{equation}
where the code blocks are illustrated
as though for a three-qubit code for simplicity.
In order to extract reliable error information,
the syndrome extraction is repeated for some times.
An optimized way to extract the syndrome information
was also proposed in Ref. \cite{Plenio97},
where the subsequent syndrome extraction is conditionally performed
according to the preceding syndrome information.
Another interesting QEC gadget based on teleportation
was proposed by Knill \cite{Knill05a}, which is illustrated as follows:
\begin{equation}
\scalebox{.35}{\includegraphics*[0cm,0.5cm][20cm,11cm]{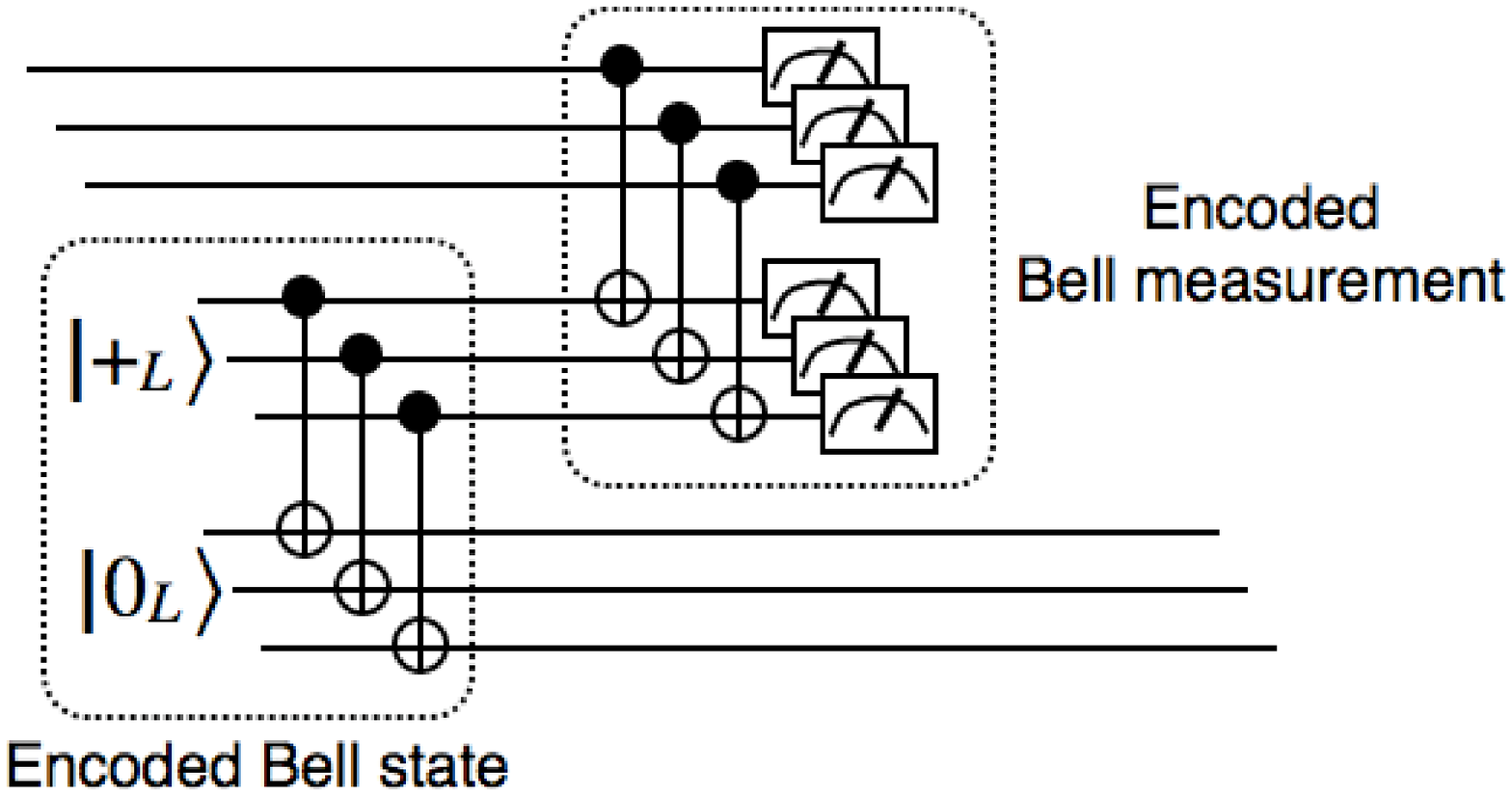}}
\label{eqn:QEC2}
\end{equation}
Here, the encoded data qubit is teleported
to the fresh encoded qubit of the ancilla Bell state.
The outcome of the encoded Bell measurement to complete the teleportation
provides sufficiently the syndrome information,
namely error-detecting or error-correcting teleportation.
Thus, it is not necessary to repeat the syndrome extraction
in this QEC gadget.
The outcome of the Bell measurement is properly propagated
to the subsequent computation as the Pauli frame
\cite{Knill05a,Niel06}.

Concatenated computation with QEC gadgets
can be employed to achieve high accuracy for logical gate operations.
In the usual fault-tolerant architectures
based on the circuit-model computation \cite{Kitaev97,Knill98,AB-O99},
the concatenation is implemented
by replacing a physical (lower-level) gate operation
recursively with a logical (upper-level) one followed by QEC gadgets
such as the circuits (\ref{eqn:QEC1}) and (\ref{eqn:QEC2}).
It is illustrated for a controlled-NOT (CNOT) gate as follows:
\begin{equation}
\scalebox{.35}{\includegraphics*[0cm,0.5cm][17cm,7.5cm]
{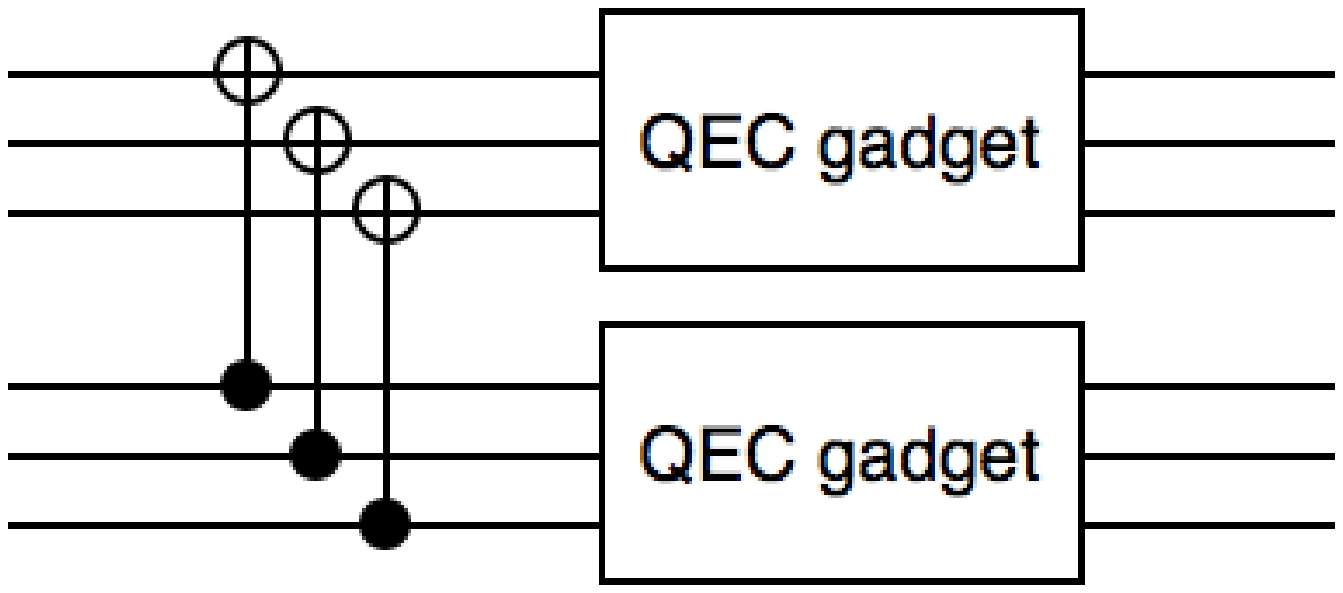}}
\end{equation}
Here, we note that any logical gate operation
should be followed by the QEC gadgets for fault-tolerant computation.
We may call this type of concatenation in terms of logical circuits
the circuit-based concatenation
or circuit-based fault-tolerant architecture.

\section{Cluster-based fault-tolerant architecture}
\label{sec:simple model}

\subsection{Main concept}

The cluster-based architecture
pursues logical cluster states with high fidelity for reliable computation,
whereas the circuit-based architecture
concerns logical circuits with high accuracy
as described in the preceding section.
(Here, the terms ``circuit-based" and ``cluster-based"
refer to the type of fault-tolerant concatenation.
They do not specify the physical-level computation.)
In the cluster model, quantum computation is implemented
through measurements of the logical qubits in cluster states.
Thus, high fidelity cluster states
directly mean the ability to perform quantum computation with high accuracy.
It is, however, not a trivial task to prepare such large entangled states
with high fidelity as cluster states of logical qubits encoded
in a concatenated QEC code.
This may be done by adopting postselection
(or multi-partite entanglement purification).
That is, logical cluster states are constructed
through verification process;
they are discarded if infection of errors is found.
It is expected generally
that as the size of an entangled state gets large,
the probability to pass the postselection decreases substantially.
Thus, we have to design suitably the cluster-based architecture
so as to make it scalable, while the postselection is made successfully.
This dilemma between postselection and scalability in concatenation
can be overcome by utilizing
the unique feature of the cluster-model computation \cite{FY09}.
The key elements are as follows:

$ \bullet $ {\bf Fundamental clusters} with certain topologies,
which are used to compose a whole cluster state
to implement a desired computation.

$ \bullet $ {\bf Verification protocols}, as parts of cluster states,
to postselect the successful one-way computation
for the construction of fundamental clusters.

$ \bullet $ {\bf Transversal bare C$Z$ gates} without verification,
which are used to connect the fundamental clusters deterministically
to construct the whole cluster state scalably.

We need not verify the whole of a cluster state by postselection,
which would have resulted in divergence of resources
due to the diminishing success probability.
Instead, at each concatenation level
we divide the whole cluster state (one-way computation)
into the fundamental clusters (gate operations and ancillae).
The fundamental clusters are deterministically
connected by the bare C$Z$ gates which operate transversally
on a suitable code such as the Steane seven-qubit code.
As a result, the postselection is localized within each fundamental cluster,
which reduces the resource usages dramatically,
though maintaining fault-tolerance of computation.

\subsection{Preliminary model}

We consider preliminarily a simple model
to illustrate the cluster-based architecture.
At the same time, we introduce cluster diagrams,
which are designed to describe properly the architecture.

We take one fundamental cluster as as follows:
\begin{equation}
\scalebox{.35}{\includegraphics*[0cm,0.5cm][10cm,5.5cm]{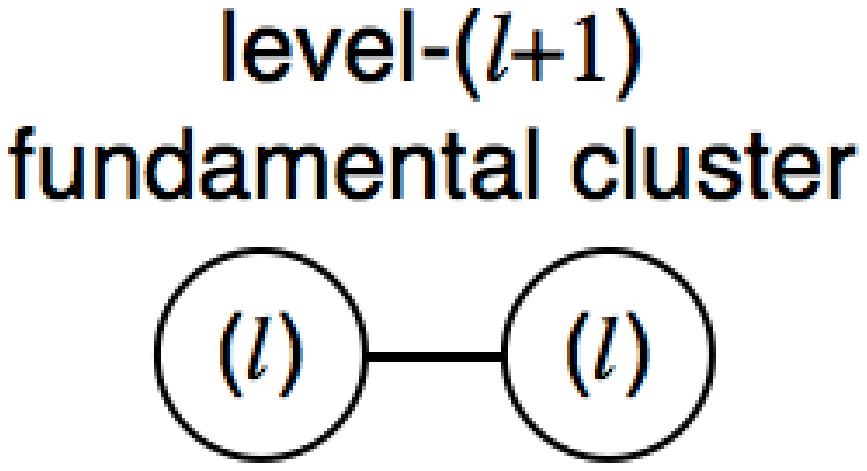}}
\label{fundamental1}
\end{equation}
Henceforth we suitably define level-$(l+1)$ fundamental clusters
as cluster states of level-$l$ qubits in concatenation of a QEC code.
(Level-0 qubits are physical ones.)
In this model the level-$(l+1)$ fundamental cluster (\ref{fundamental1})
consists of two level-$l$ qubits connected with a C$Z$ gate.
We construct this level-$(l+1)$ fundamental cluster
through a verification protocol as given in the following circuit:
\begin{equation}
\scalebox{.4}{\includegraphics*[0cm,0.5cm][17cm,8.5cm]{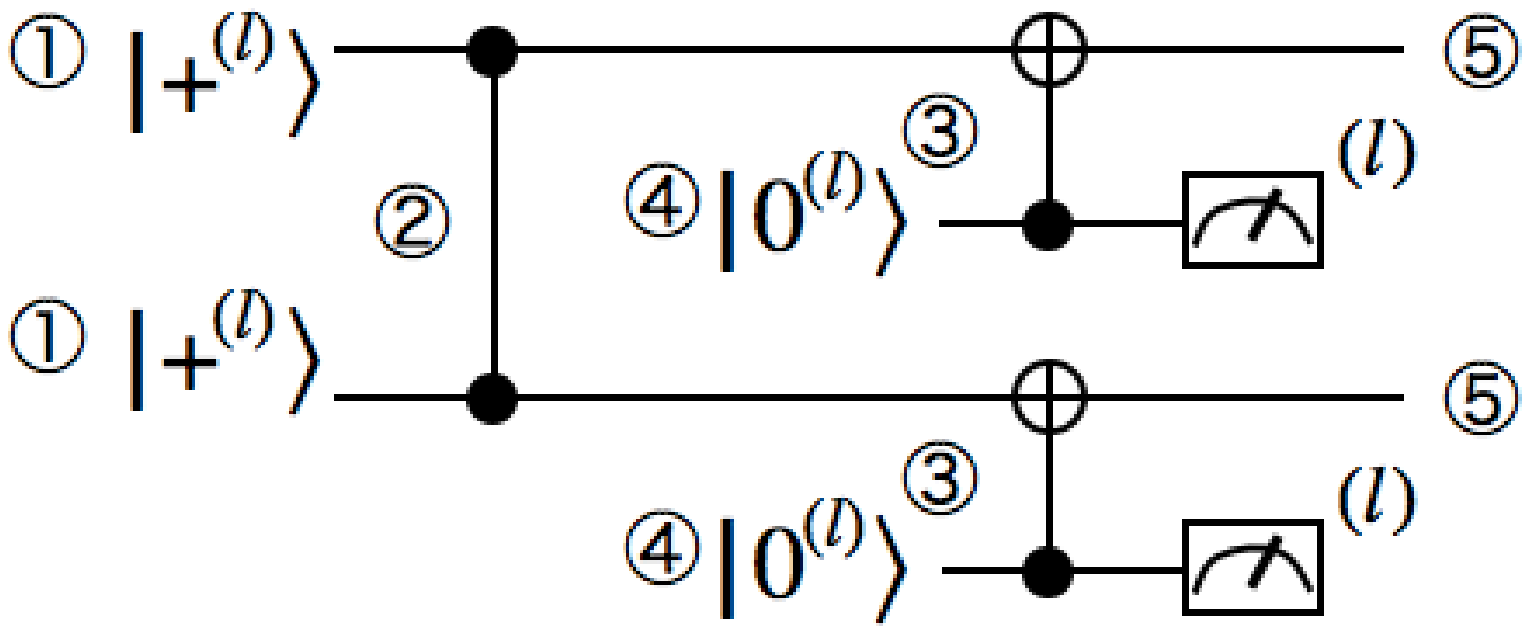}}
\label{ex-circuit}
\end{equation}
The two-qubit cluster is formed
from the two level-$l$ logical $|+^{(l)}\rangle$ qubits
(\textcircled{\small{1}})
through the C$Z$ gate operation (\textcircled{\small{2}}).
The errors which are introduced to these two qubits
before and during the C$Z$ gate operation are detected
by using a sort of the Steane's QEC gadget (\textcircled{\small{3}})
with the ancilla $|0^{(l)}\rangle$ qubits (\textcircled{\small{4}}).
This verification protocol is implemented with postselection
to obtain the level-$(l+1)$ fundamental cluster (\ref{fundamental1})
with higher fidelity (\textcircled{\small{5}}).

In the cluster-based architecture,
the entanglement operation with verification
to construct the level-$(l+1)$ fundamental cluster
is implemented by one-way computation on a certain cluster state
which is made by combining the level-$l$ fundamental clusters
with the transversal bare C$Z$ gates.
Specifically, the process (\ref{ex-circuit})
to obtain the fundamental cluster (\ref{fundamental1})
is described in terms of a cluster diagram as follows:
\begin{equation}
\scalebox{.4}{\includegraphics*[0cm,0.5cm][24cm,10cm]{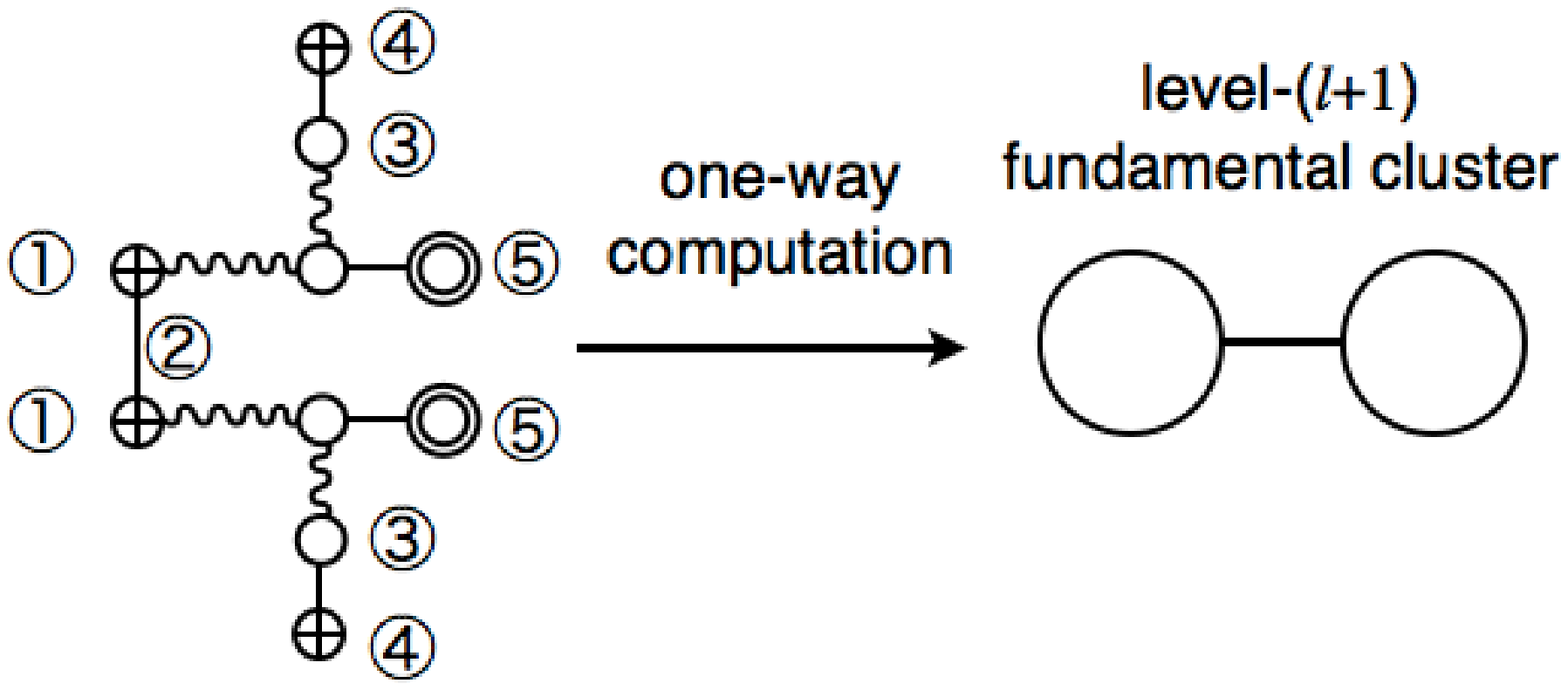}}
\label{ex-cluster}
\end{equation}
Here, the elements corresponding to those in the circuit (\ref{ex-circuit})
are labeled the same numbers
\textcircled{\small 1} -- \textcircled{\small 5}.
We occasionally use the two-dimensional diagrams such as (\ref{ex-cluster})
to abbreviate the three-dimensional arrays
to represent the whole cluster states
by omitting the coordinate for the code blocks
according to the encoding rules as explained below.
[The whole three-dimensional array of (\ref{ex-cluster})
will be illustrated later.]
The wavy lines in the diagram (\ref{ex-cluster})
indicate the bare C$Z$ gates acting transversally
on the level-$(l-1)$ qubits
composing the level-$l$ fundamental clusters.
The output qubits (\textcircled{\small 5}) are denoted by $\circledcirc$
as the verified level-$(l+1)$ fundamental cluster.
The operation for encoding and transferring
the level-$l$ code state $|+^{(l)}\rangle$ is described
by $\oplus$ symbolically:
\begin{equation}
\scalebox{.3}{\includegraphics*[0cm,0.5cm][17cm,13.5cm]{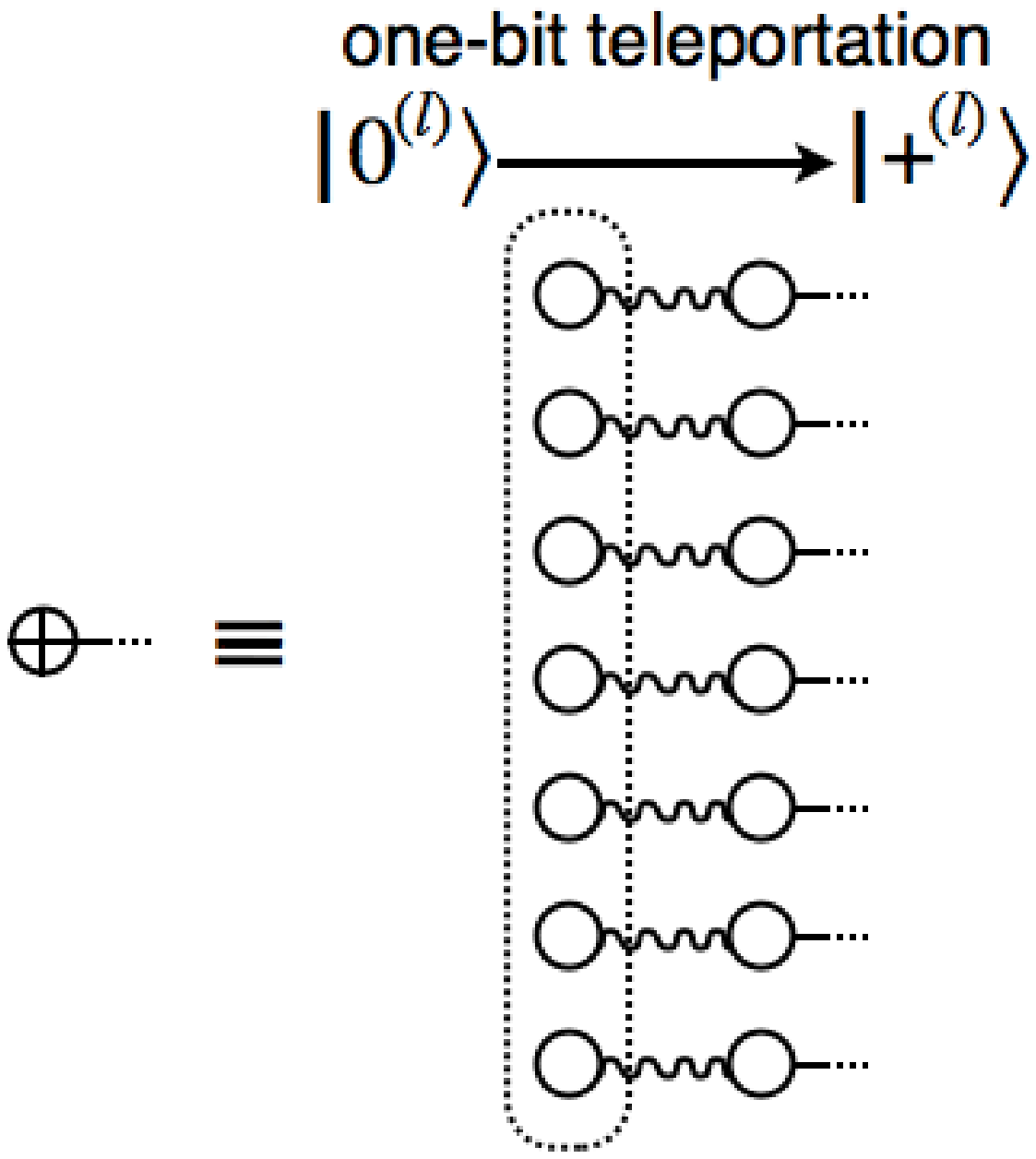}}
\label{transferplus}
\end{equation}
Here, the level-$(l-1)$ qubits surrounded by the dotted line form
the level-$l$ code state (cluster) $|0^{(l)}\rangle$.
They are teleported upon measurements to another block of qubits
as $|+^{(l)}\rangle$ by a Hadamard operation
$|+^{(l)}\rangle=H|0^{(l)}\rangle$ with bare C$Z$ gates
(one-bit teleportation).
The encoding operation of the level-$l$ code state $|0^{(l)}\rangle$
is also described by {\LARGE $ \bullet $} symbolically
for the later use:
\begin{equation}
\scalebox{.3}{\includegraphics*[0cm,0.5cm][17cm,13.5cm]{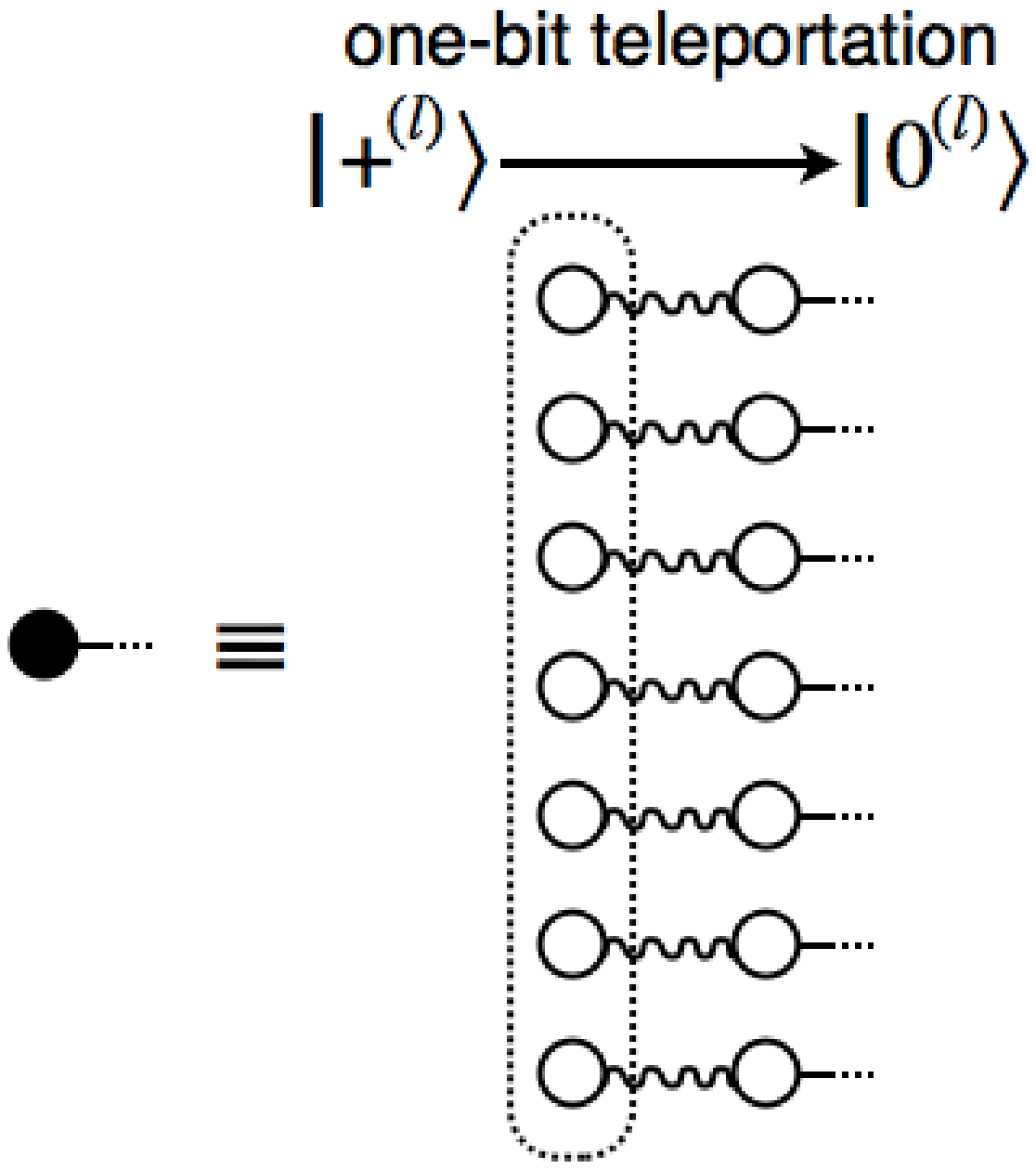}}
\label{transferzero}
\end{equation}
By applying the $\oplus$ encoding (\ref{transferplus}),
the full three-dimensional array of the diagram (\ref{ex-cluster})
is obtained with the axes corresponding to the code blocks, logical qubits
and time as follows:
\begin{equation}
\scalebox{.3}{\includegraphics*[0cm,0.5cm][28cm,19cm]{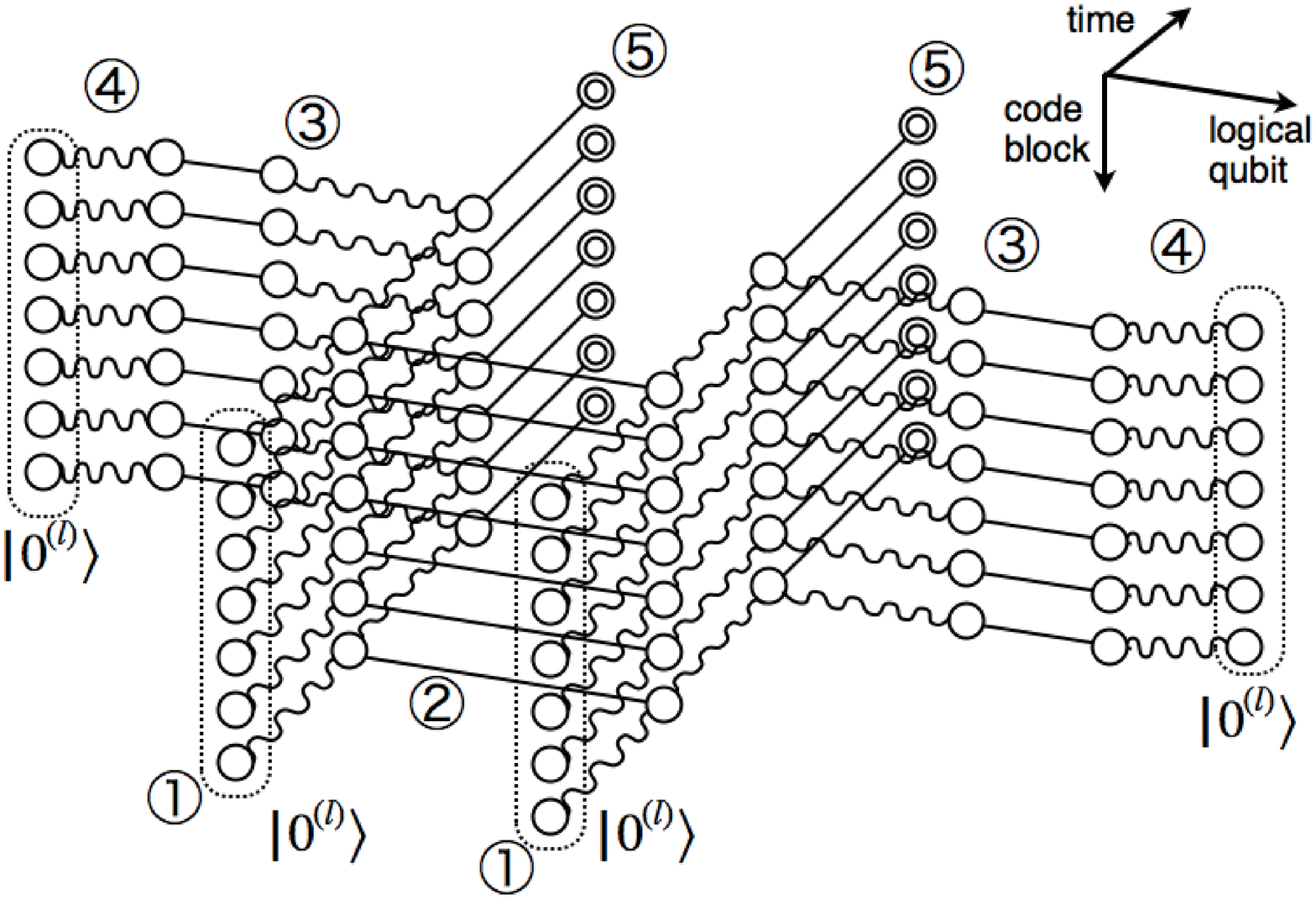}}
\label{example-full}
\end{equation}
Here, we observe that the level-$(l+1)$ fundamental cluster
(\ref{fundamental1}) is constructed through the verification
by using $ 5 \times 7 $ level-$l$ fundamental clusters (\ref{fundamental1})
and 4 level-$l$ logical qubits $|0^{(l)}\rangle$
which are suitably connected with $ (4+4) \times 7 $
level-$(l-1)$ bare C$Z$ gates.

As seen in the diagrams (\ref{transferplus}) and (\ref{transferzero}),
the level-$l$ code states $|0^{(l)}\rangle$ and $|+^{(l)}\rangle$
are used for the encoding operations.
They are given as the cluster states of level-$(l-1)$ qubits,
which are similar to the fundamental cluster (\ref{fundamental1}).
[See the diagrams (\ref{logical-qubit-zero})
and (\ref{logical-qubit-plus}) in the next section.]
We can prepare these cluster states
for $|0^{(l)}\rangle$ and $|+^{(l)}\rangle$
by combining some copies of the level-$l$ fundamental cluster
(\ref{fundamental1}) with the level-$(l-1) $ bare C$Z$ gates.
(Here, we do not present their preparation explicitly
for this preliminary model.)
An alternative option is to include
the level-$l$ code states $|0^{(l)}\rangle$ and $|+^{(l)}\rangle$
in the set of level-$l$ fundament clusters,
as will be adopted in the next section
for an efficient construction of fundamental clusters with high fidelity.

The circuit (\ref{ex-circuit}) is executed
in the diagram (\ref{ex-cluster}) or (\ref{example-full})
by measuring the the level-$(l-1)$ qubits
except for the output $\circledcirc$ qubits.
Then, the syndrome information is extracted from the measurement outcomes
(\textcircled{\small{3}} and \textcircled{\small{4}}).
If this level-$l$ syndrome is found to be correct,
we keep the output $\circledcirc$ qubits (\textcircled{\small{5}})
as the verified level-$(l+1)$ fundamental cluster.
Otherwise, we discard the unsuccessful outputs.
This one-way computation completes one concatenation;
the level-$(l+1)$ fundamental cluster
as the entangled set of output level-$l$ qubits ($\circledcirc$'s)
has been constructed and verified
by using the level-$l$ fundamental clusters with bare C$Z$ gates.

We produce many copies of the fundamental cluster
by performing the above procedure recursively
up to a certain logical level high enough to achieve the expected accuracy.
Then, we construct the whole cluster state
to implement a desired computation
by combining these copies of the fundamental cluster
with the transversal bare C$Z$ gates.
The logical error of the transversal bare C$Z$ gate on the concatenated code
also becomes sufficiently small at the highest level.
Thus, given the clean enough fundamental clusters at the highest level,
the one-way computation is operated fault-tolerantly
on the whole cluster state.
In this preliminary model, however, the noise threshold will be rather low,
since the verification protocol is not optimal,
and some of the qubits are connected doubly to the bare C$Z$ gates.
A more efficient architecture will be described in the next section,
which achieves a high noise threshold $ \sim 3 \% $.

\subsection{Unique features}

We should mention that the role of bare C$Z$ gates
in the cluster-based architecture
provides the essential distinction from the circuit-based architecture.
The postselection with QEC gadgets
can really achieve high accuracy for computation.
However, in the circuit-based concatenation
the postselection of gate operations should be performed
in the ongoing computation
(even if the error-detecting teleportation is utilized
with off-line preparation of ancilla states \cite{Knill05a}).
Thus, if errors are detected, the computation should be restarted
from the beginning, which results in divergence of resources usage.
This is because in the circuit-based architecture
any logical gate operation is necessarily followed by QEC gadgets
at each concatenation level, as seen in Sec. \ref{sec:back ground}.

Instead, in the cluster-based architecture
bare C$Z$ gates, which are not accompanied by QEC gadgets,
are partially used for the one-way computation
to implement the construction process,
while fault-tolerance can be ensured
by the verification and postselection of fundamental clusters.
The logical cluster states are really postselected
{\it off-line and locally}
since the whole cluster is divided into the fundamental clusters
with the help of bare C$Z$ gates.
When clean enough fundamental clusters are just constructed,
we connect them with bare C$Z$ gates deterministically,
and then start the computation.
The fundamental clusters, which represent the gate operations,
have been constructed successfully in advance
by removing sufficiently the errors
via the postselection in the lower-level one-way computation,
before starting the computation at the higher level.
Thus, we may call this verification process
as {\it preselection} or {\it error-precorrection} of gate operations.
Here, it should be noted
that the postselection for the whole cluster state or computation,
without the use of bare C$Z$ gates, increases exponentially the resources
according to the computation size.
The present architecture certainly succeeds
in reconciling postselection with scalability,
which would not be achieved without the cluster-model computation.

The cluster-based architecture also exploits a good transversal property
by adopting a suitable code such as the Steane seven-qubit code.
That is, the operations on the physical qubits
are all transversal, and really limited
after the verification process at the lowest (physical) level.
In fact, as seen in the diagram (\ref{ex-cluster}),
any direct operation is not implemented on the output qubits
($\circledcirc$'s)
through the verified construction of fundamental cluster.
The desired entanglement among them to form the fundamental cluster
at the next level is rather generated
via one-bit teleportation in the one-way computation.
Thus, they inherit transversally
the errors on the constituent physical-level qubits,
up to the Pauli frame information
from the one-way computation for cluster construction.
Then, these output qubits composing the fundamental clusters
undergo the transversal bare C$Z$ gates and measurements
at the next level for the first time.
This transversal property provides a simple structure
of logical errors in concatenation
to estimate readily the noise threshold.
In this respect, the cluster-based architecture
presents a practical way to construct large entangled states,
including the concatenated code states and fundamental clusters,
the errors of which are described in a good approximation
by the homogeneous errors on the constituent physical-level qubits
\cite{Eastin07}.
The details will be demonstrated in the following sections.

\section{Concatenated construction of verified cluster states}
\label{sec:concatenated construction}

We now introduce an efficient architecture for fault-tolerant concatenation
by adopting a set of suitable fundamental clusters
and elaborate verification protocols.
It is really designed to achieve high noise threshold
by taking full advantage of the present cluster-based scheme.
As seen in the diagram (\ref{ex-cluster}),
some of the qubits are connected doubly to the bare C$Z$ gates
for the cluster construction in the preliminary model.
This lowers the noise threshold substantially.
Thus, the topologies of the fundamental clusters
should be chosen so as to limit suitably the use of bare C$Z$ connections
(at most one bare C$Z$ gate to each qubit)
and redundant qubits for the cluster construction.
It should also be noted
that the errors on the resultant fundamental clusters
are not detected after the construction is completed.
This requires that the verification protocols
should detect fully the first-order errors
except for some of the errors introduced by the final few operations,
which are inevitably left on the output states.

\subsection{Fundamental clusters}

We adopt the following states as the level-$l$ fundamental clusters:
\begin{equation}
| h^{(l)} \rangle ,  | 0^{(l)} \rangle ,  | +^{(l)} \rangle .
\end{equation}
They are depicted in terms of the cluster diagrams as
\begin{equation}
\scalebox{.25}{\includegraphics*[0cm,0.5cm][20cm,6cm]{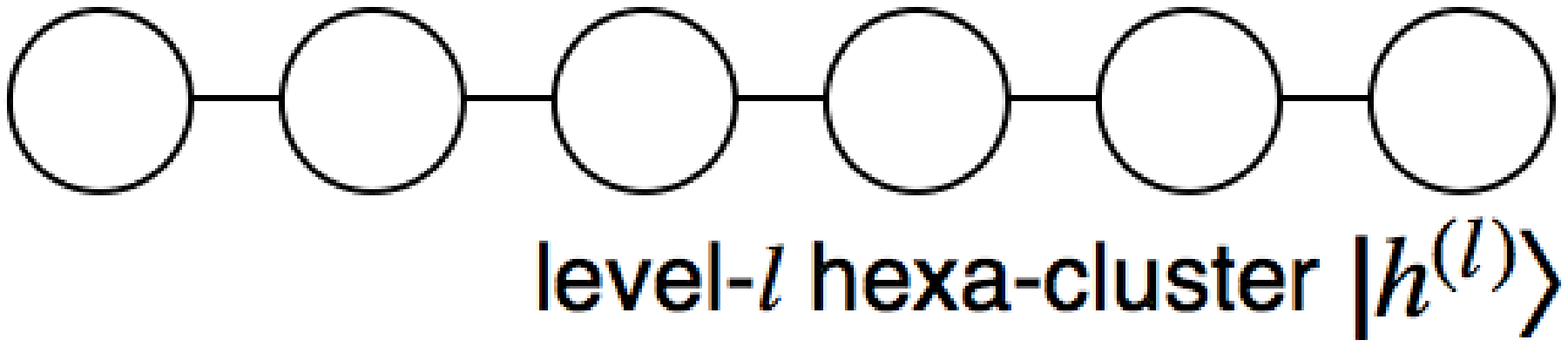}}
\label{hexadefinition}
\end{equation}
\begin{equation}
\scalebox{.3}{\includegraphics*[0cm,0.5cm][14cm,10.5cm]
{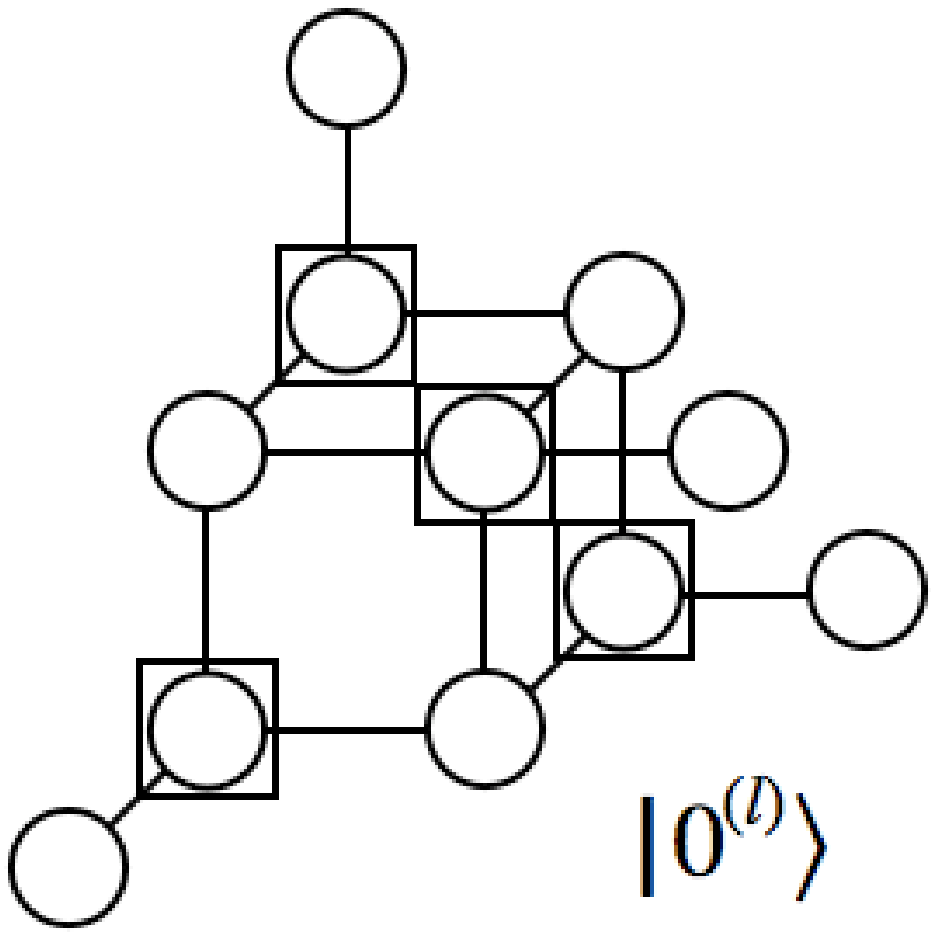}}
\label{logical-qubit-zero}
\end{equation}
\begin{equation}
\scalebox{.3}{\includegraphics*[0cm,0.5cm][14cm,10.5cm]
{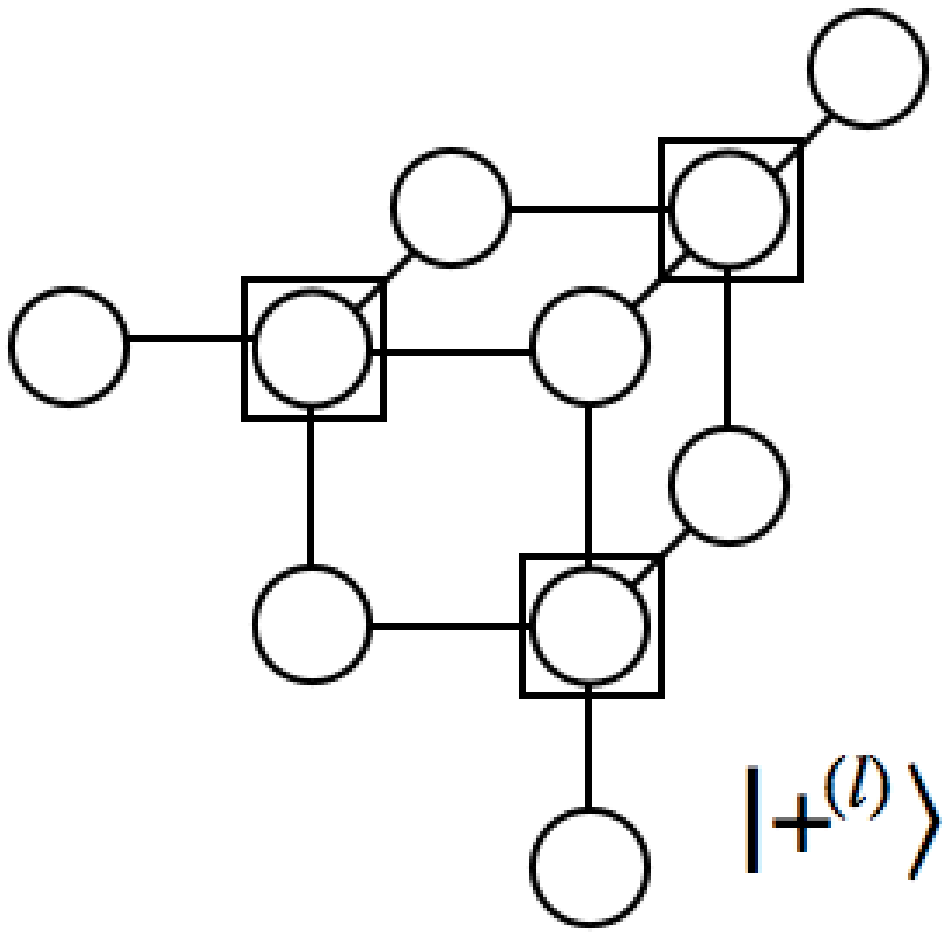}}
\label{logical-qubit-plus}
\end{equation}
where the circles denote the level-$(l-1)$ qubits,
and the boxed qubits are measured for Hadamard operations
to obtain $ | 0^{(l)} \rangle $ and  $ | +^{(l)} \rangle $.
The hexa-cluster $ | h^{(l)} \rangle $
is a cluster state of six level-$(l-1)$ qubits
which are connected linearly with C$Z$ gates.
This hexa-cluster represents an elementary unit of gate operations
as seen later.
The level-$l$ concatenated code states
$ | 0^{(l)} \rangle $ and $| +^{(l)} \rangle$
are also taken as the fundamental clusters in this architecture.
They are used as ancillae for encoding and syndrome detection.

\subsection{Single and double verifications}

The level-$(l+1)$ fundamental clusters are constructed
by operating the C$Z$ gates on the level-$l$ qubits.
These gate operations inevitably introduce errors on the output states.
Thus, as seen in Sec. \ref{sec:simple model},
we verify and postselect the output states
for the high fidelity construction.
Specifically, we detect the errors efficiently
by combining two verification gadgets, namely
single and double verifications.

The C$Z$ gate operation with single verification is given
in terms of a circuit as
\begin{equation}
\scalebox{.35}{\includegraphics*[0cm,1cm][15cm,8cm]{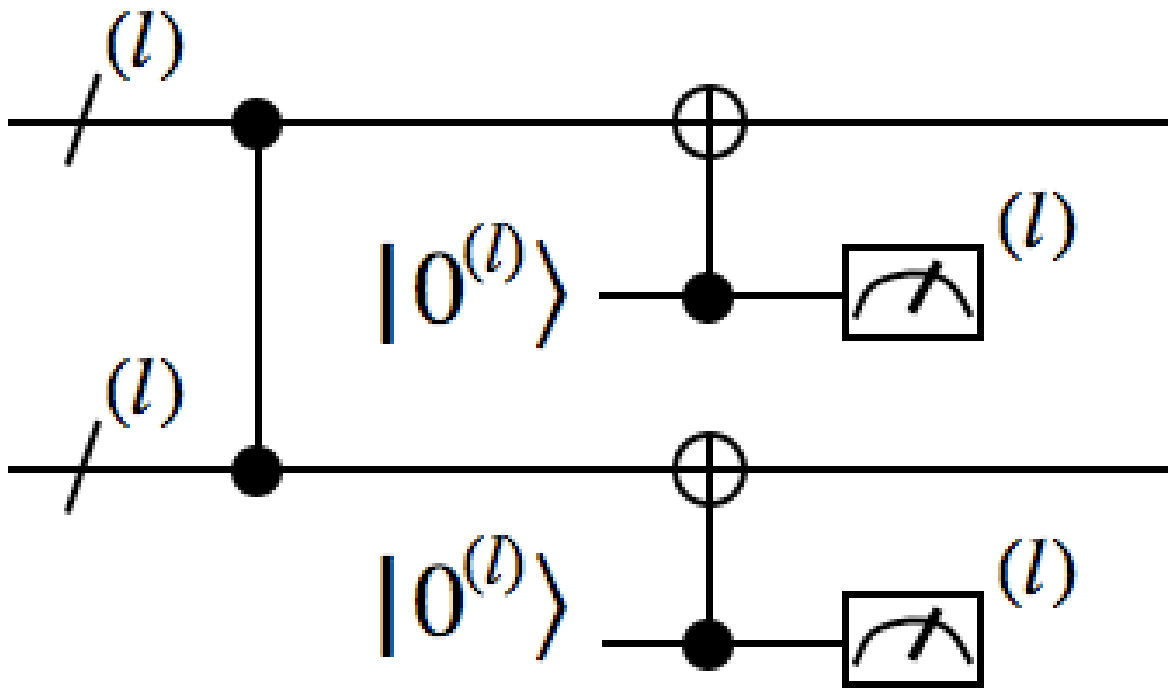}}
\label{single_circuit}
\end{equation}
where each dashed line with index $(l)$ indicates
that seven level-$(l-1)$ wires are contained there.
The single verification is the same as the protocol (\ref{ex-circuit})
for the model in Sec. \ref{sec:simple model}.
The $Z$ error on the level-$l$ qubit is detected
by the $Z$ syndrome extraction after the C$Z$ gate operation.
Furthermore, the preceding $ X $ error on the level-$l$ qubit
is detected by the $Z$ syndrome extraction for the other level-$l$ qubit
since it is propagated through the C$Z$ gate as a $Z$ error.

The cluster diagram for the single verification (\ref{single_circuit})
is given with the fundamental clusters as
\begin{equation}
\scalebox{.35}{\includegraphics*[0cm,1cm][13cm,9cm]{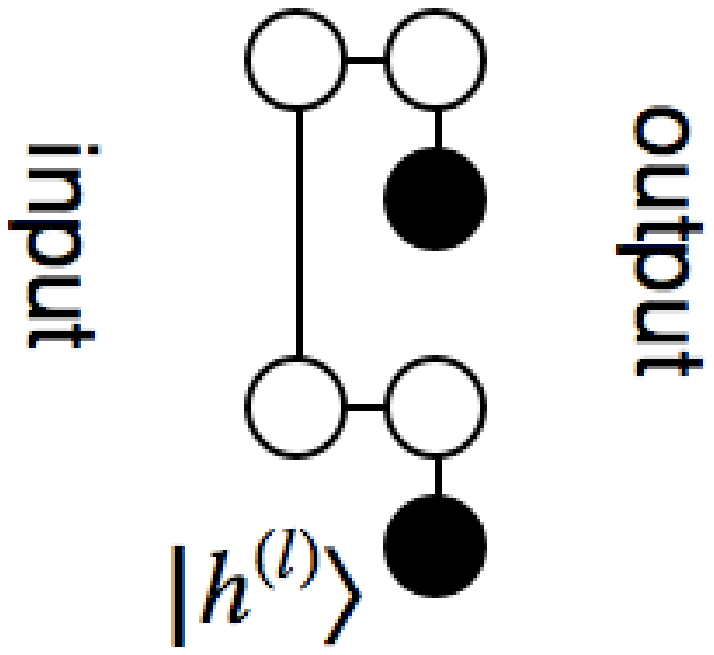}}
\label{single-cluster}
\end{equation}
where the {\LARGE $ \bullet $}'s denote the encoding of $|0^{(l)}\rangle$
for the syndrome extraction in the circuit (\ref{single_circuit}).
By considering the {\LARGE $ \bullet $} encoding (\ref{transferzero}),
the single-verification diagram (\ref{single-cluster})
is fully illustrated in terms of a cluster state of level-$(l-1)$ qubits as
\begin{equation}
\scalebox{.25}{\includegraphics*[0cm,0.5cm][25cm,20cm]{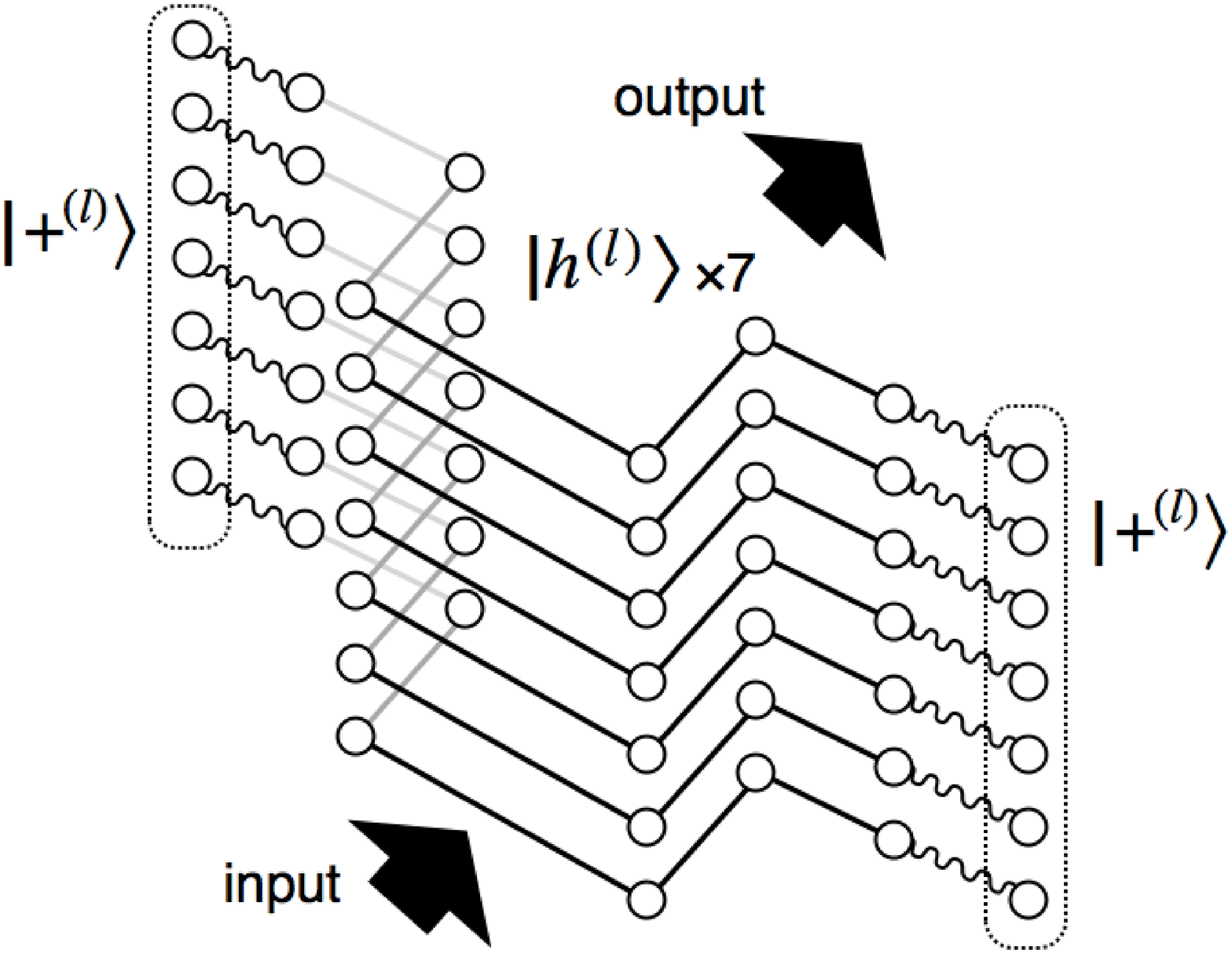}}
\label{single-verification-full}
\end{equation}
which may be compared with the diagram (\ref{example-full})
in the preliminary model.
Here, we observe that
the level-$l$ C$Z$ gate operation with single verification,
as given in the circuit (\ref{single_circuit})
and diagram (\ref{single-cluster}),
is implemented by using $ 7 | h^{(l)} \rangle $'s,
$ 2 | +^{(l)} \rangle $'s and $ 2 \times 7 $ level-$(l-1)$ bare C$Z$ gates.

In order to remove sufficiently the errors
in the final stage of construction,
we implement the double verification,
which may be viewed as a sophistication of the Steane's QEC gadget
in the circuit (\ref{eqn:QEC1}).
The C$Z$ gate operation with double verification is described as follows:
\begin{equation}
\scalebox{.4}{\includegraphics*[0cm,1cm][19cm,10cm]{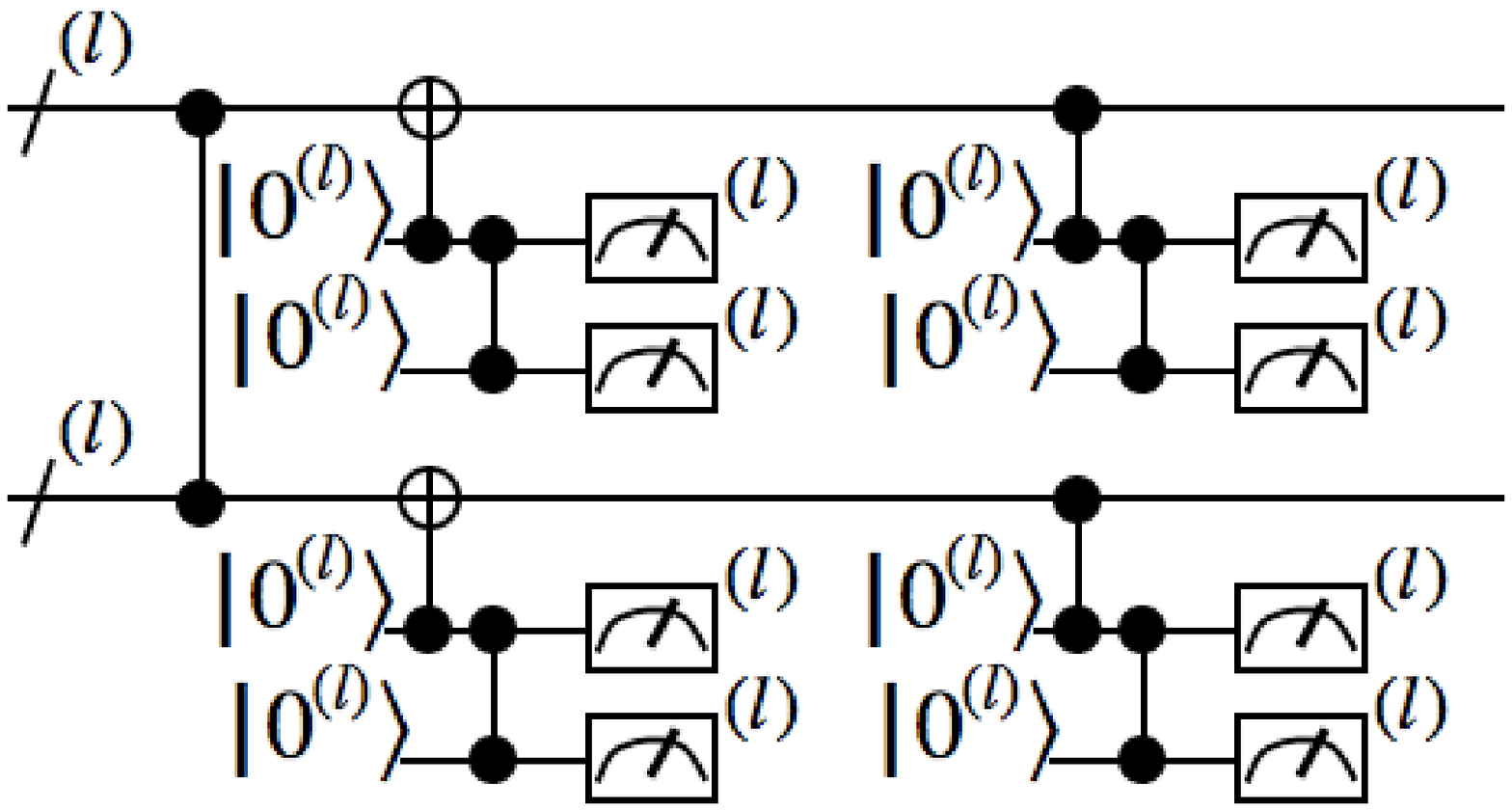}}
\label{double_circuit}
\end{equation}
Here, the $Z$ error verification through a CNOT gate
is followed by the $X$ error verification through a C$Z$ gate
for high fidelity.  Furthermore, the error propagation
from the primary ancilla qubit $|0^{(l)}\rangle$
to the data qubit through the two-qubit gate (CNOT or C$Z$) is prohibited
in the leading order by inspecting the primary $|0^{(l)}\rangle$
with the secondary $|0^{(l)}\rangle$.
In fact, this double verification
with the primary and secondary ancilla states has been applied recently
to implement a high-performance recurrence protocol
for entanglement purification \cite{FY09EP},
where its optimality for detecting the first-order errors is discussed.
We also note that the single and double verifications
in (\ref{single_circuit}) and (\ref{double_circuit})
both remove the preceding errors through the C$Z$ gate
by the syndrome extractions for the two level-$l$ qubits.

Similar to the single-verification diagram (\ref{single-cluster}),
the circuit (\ref{double_circuit}) for the double verification
is implemented by a cluster diagram as follows:
\begin{equation}
\scalebox{.3}{\includegraphics*[0cm,1cm][17cm,10cm]{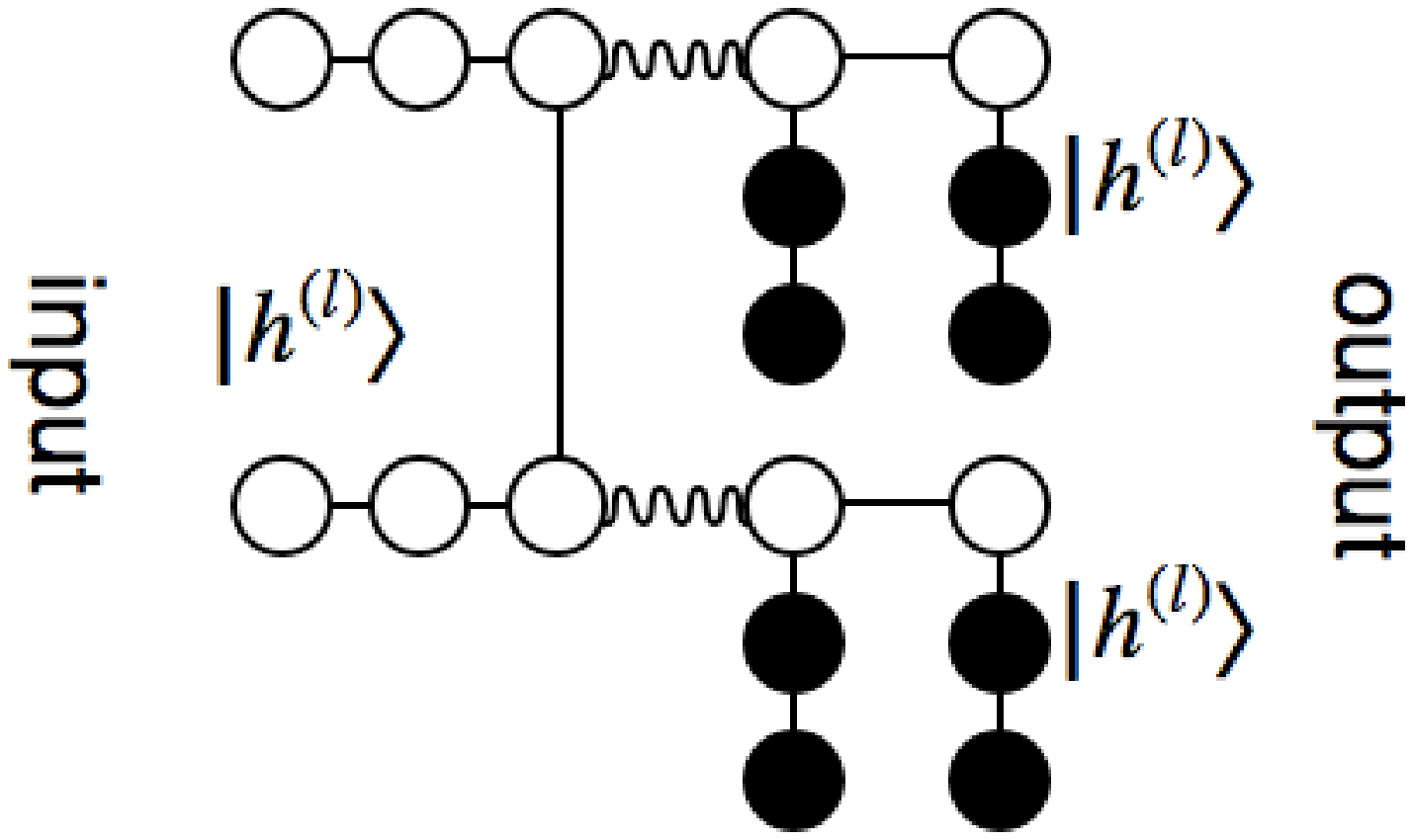}}
\label{double-cluster}
\end{equation}
The full diagram for (\ref{double-cluster})
is generated by considering the {\LARGE $ \bullet $} encoding
of $|0^{(l)}\rangle$ in (\ref{transferzero}),
similarly to the single-verification diagram
(\ref{single-verification-full}).
We realize in the diagram (\ref{double-cluster})
that the level-$l$ C$Z$ gate operation
with double verification is implemented
by combining $ 3 \times 7 $ $ | h^{(l)} \rangle $'s
and $ 8 | +^{(l)} \rangle $'s
with $ (8+2) \times 7 $ level-$(l-1)$ bare C$Z$ gates.

\subsection{Concatenated cluster construction}

The level-$(l+1)$ fundamental clusters
are constructed from the level-$l$ ones via one-way computation.
In order to achieve high fidelity,
the C$Z$ gate operations with single and double verifications
are combined by using the bare C$Z$ gates in a suitable way;
(i) each qubit has at most one bare C$Z$ connection (wavy line),
and (ii) the output {\large $ \circledcirc $} qubits
to form the level-$(l+1)$ fundamental clusters
have no bare C$Z$ connection,
and they are doubly verified in the final stage of construction.
Specifically, the level-$(l+1)$ hexa-cluster $ | h^{(l+1)} \rangle $
is constructed as follows:
\begin{equation}
\scalebox{.3}{\includegraphics*[0cm,0cm][25cm,16cm]{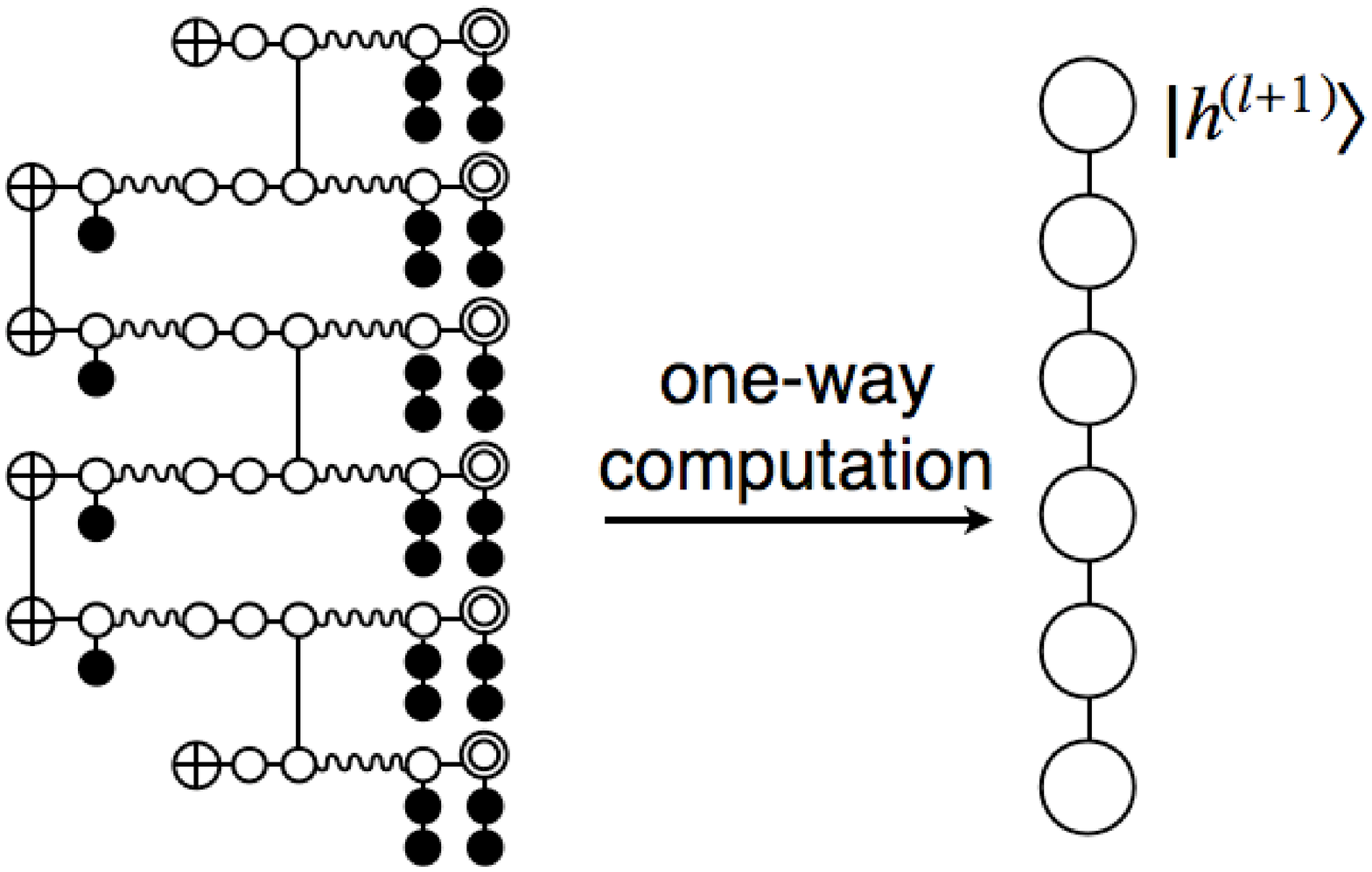}}
\label{hexa}
\end{equation}
The 6 $ | +^{(l)} \rangle $'s are transferred
by the $ \oplus $ encoding (\ref{transferplus}),
and they are entangled through 2 C$Z$ gates with single verification
(\ref{single-cluster})
and 3 C$Z$ gates with double verification (\ref{double-cluster})
to form the $ | h^{(l+1)} \rangle $
(the output 6 {\large $ \circledcirc $} qubits at the level $ l $).
This one-way computation to construct the $ | h^{(l+1)} \rangle $
is implemented by measuring the level-$ (l-1) $ qubits,
except those for the output {\large $ \circledcirc $}'s,
in the three-dimensional diagram for (\ref{hexa}).
[The full diagram is generated with the code-block axis
supplemented according to the encodings
(\ref{transferplus}) and (\ref{transferzero}),
as the diagrams (\ref{example-full}) and (\ref{single-verification-full}).]
The level-$ l $ syndromes are extracted through the measurements
of the ancilla encoded {\LARGE $ \bullet $} qubits.
If all the level-$ l $ syndromes are correct,
the entangled set of six level-$l$ {\large $ \circledcirc $} qubits
survive as a verified $ | h^{(l+1)} \rangle $.

Since the cluster diagrams such as (\ref{hexa}) look somewhat complicated,
we introduce suitably the reduced diagrams
by omitting the time axis and qubits measured in the one-way computation.
The hexa-cluster construction (\ref{hexa}) is described as follows:
\begin{equation}
\scalebox{.3}{\includegraphics*[0cm,0.7cm][20cm,4cm]{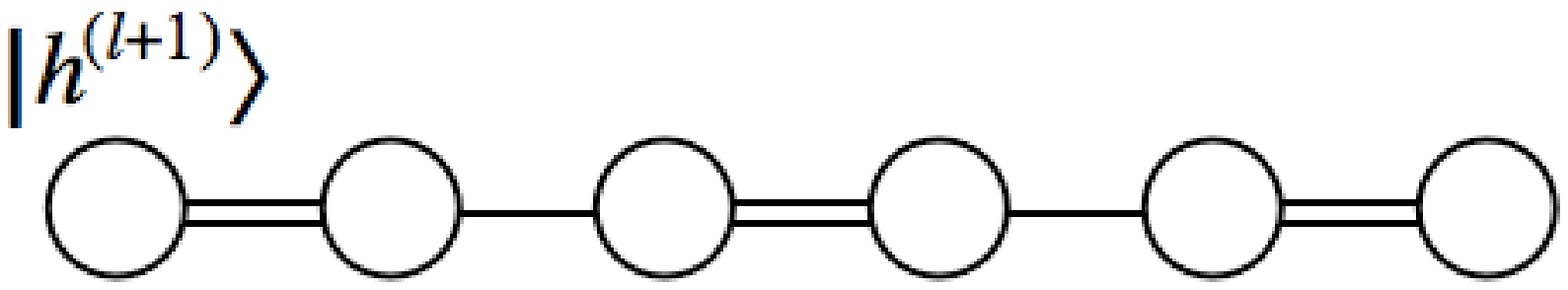}}
\label{hexa-reduced}
\end{equation}
Here, the single and double lines indicate
the single and double verifications, respectively,
and it is understood that
the single verifications are always done before the double verifications.
We construct similarly the fundamental clusters
$|0^{(l+1)}\rangle$ and $|+^{(l+1)}\rangle$ as
\begin{equation}
\scalebox{.3}{\includegraphics*[0cm,0.5cm][22cm,11cm]{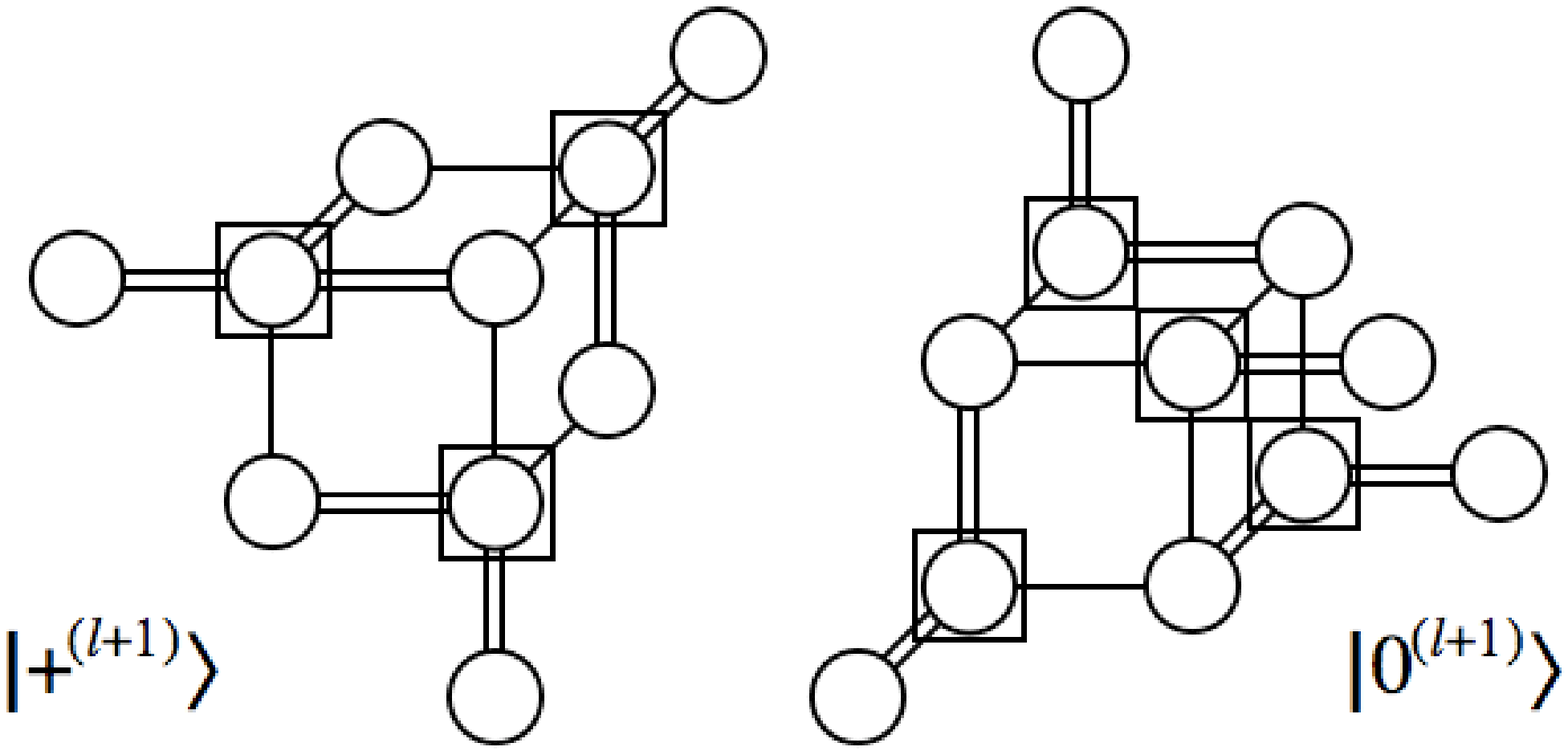}}
\label{qubit-reduced}
\end{equation}
where the boxed level-$l$ qubits are measured transversally
in the $X$ basis for Hadamard operations.
We see that in these reduced diagrams
all the qubits have at least one double-line connection,
that is they are doubly verified in the final stage of construction.
We can produce systematically the construction processes
such as (\ref{hexa}) from the reduced diagrams.
The details are described in the Appendix \ref{sec:code state}.

At the beginning of concatenation,
the construction of the level-2 fundamental clusters
by the physical-level computation
is somewhat different from the constructions at the higher levels.
This is because the verified level-1 fundamental clusters
are not available by definition from the lower-level construction.
It may be suitable to adopt the circuit-model computation
at the physical level since both CNOT and C$Z$ gates
are deterministically available.
The level-1 $ | 0^{(1)} \rangle $ and $ | +^{(1)} \rangle $ 
are first encoded and verified against the $Z$ and $X$ errors
by measuring the $X$ and $Z$ stabilizers, respectively.
They are, however, not clean enough for the present purpose.
We secondly verify the $X$ and $Z$ errors
on the $ | 0^{(1)} \rangle $ and $ | +^{(1)} \rangle $, respectively,
as follows:
\begin{equation}
\scalebox{.4}{\includegraphics*[0cm,0.5cm][20cm,6cm]{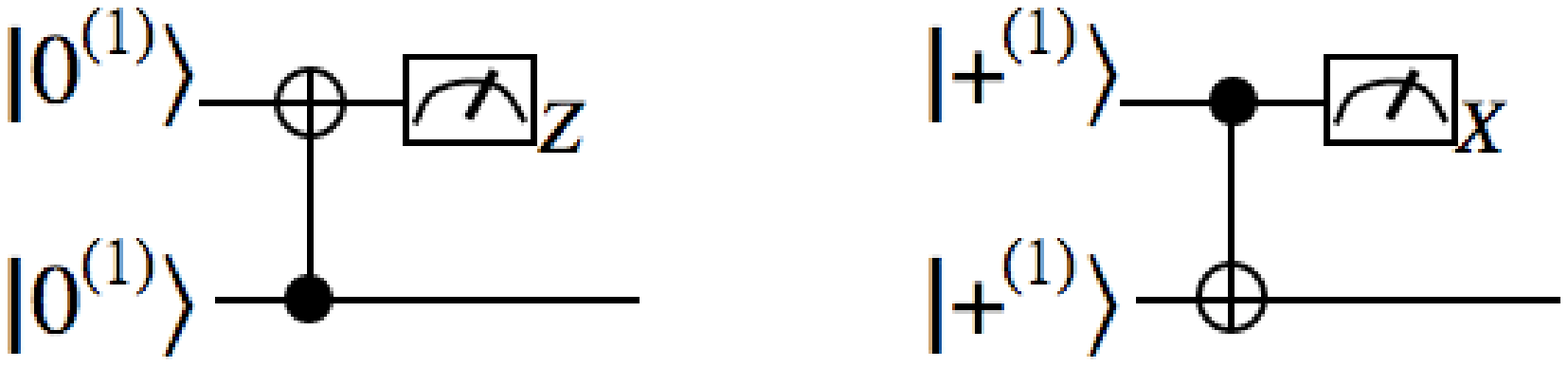}}
\label{qubit_level1}
\end{equation}
This operation is the same as the multi-partite entanglement
purification \cite{DAB03ADB05}.
Then, we construct the level-2 fundamental clusters
$ | h^{(2)} \rangle $, $ | 0^{(2)} \rangle $ and $ | +^{(2)} \rangle $
from these verified level-1 qubits
$ | 0^{(1)} \rangle $ and $ | +^{(1)} \rangle $ by implementing the circuits
(\ref{single_circuit}) and (\ref{double_circuit})
with the bare C$Z$ gates ($l=1$) according to the reduced diagrams
(\ref{hexa-reduced}) and (\ref{qubit-reduced}).
It is also possible to perform the physical-level one-way computation
by means of the cluster diagrams to implement the relevant circuits
for the level-2 construction.
Additional errors are, however, introduced
lowering slightly the noise threshold
since the extra operations are required
for the CNOT gate operations in the one-way computation.
This will be considered explicitly in Sec. \ref{sec:miscellaneous}.

\subsection{Universal computation}

The fundamental clusters are constructed through verification
up to the highest logical level $ {\bar l} $
to achieve the fidelity required for a given computation size.
Then, we can perform accurately the computation with Clifford gates
by combining the highest-level hexa-clusters
$ | h^{({\bar l}+1)} \rangle $ with the transversal bare C$Z$ gates
and performing the Pauli basis measurements of the level-$ \bar{l} $ qubits
in the cluster states.
Furthermore, we can implement even non-Clifford gates
for universal computation as explained below.

In the cluster model the operation $ H Z( \theta ) = H e^{-i \theta Z / 2} $
is implemented by the measurement in the basis
$ Z( \pm \theta ) \{ | + \rangle , | - \rangle \} $ with $ \pm \theta $
to be selected according to the outcome of preceding measurements
\cite{OWC}.
The non-Clifford gates, e.g., the $ \pi / 8 $ gate = $ Z( \pi / 4 ) $,
however, do not operate transversally even on the Steane seven-qubit code.
Then, in order to implement the $ \pi / 8 $ gate
with a transversal measurement, we make use of the equivalence as follows:
\begin{equation}
\scalebox{.3}{\includegraphics*[0cm,0.5cm][17cm,7cm]{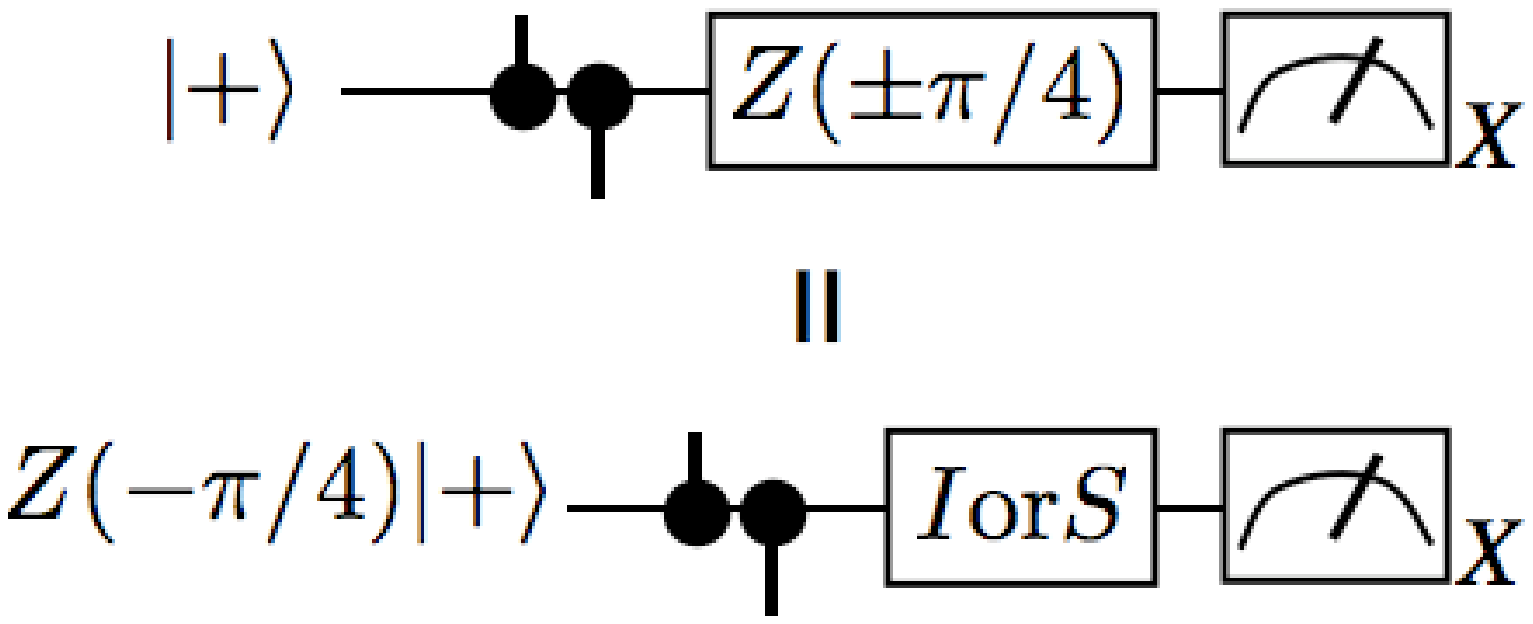}}
\label{pi8equality}
\end{equation}
As a result, the operation $ H Z( \pi / 4 ) $ can be implemented
by the preparation of the state $ Z( - \pi / 4 ) | + \rangle $
and the measurement with the $ I $ or $ S = Z( \pi / 2 ) $ operation
(the selection of measurement basis $X$ or $-Y=SXS^{\dag}$).
The preparation of $ Z( - \pi / 4 ) | + \rangle $
is reduced to that of $ | \pi / 8 \rangle
= \cos ( \pi / 8 ) | 0 \rangle + \sin ( \pi / 8 ) | 1 \rangle $
based on the relation
\begin{equation}
Z( - \pi / 4 ) | + \rangle = e^{i \phi} HS | {\pi / 8}  \rangle ,
\end{equation}
where $ \phi $ is a certain phase.
In this way we can implement the $ H $, $ S $, $ \pi / 8 $ and C$Z$ gates
as a universal set by the transversal Pauli basis measurements
of the level-$\bar{l}$ qubits,
including $ | {\pi / 8}^{({\bar l})} \rangle $,
in the level-$(\bar{l}+1)$ cluster states
\cite{FY07,Silva07}.

The level-1 $ | {\pi / 8}^{(1)} \rangle $
is encoded by the usual method \cite{Knill98,Knill05a}.
Then, similarly to the other fundamental clusters
the upper-level $ | {\pi / 8}^{(l+1)} \rangle $ ($l \geq 1$)
is encoded with the lower-level $ | {\pi / 8}^{(l)} \rangle $,
as shown in the following reduced diagram:
\begin{equation}
\scalebox{.25}{\includegraphics*[0cm,0.5cm][19cm,16cm]{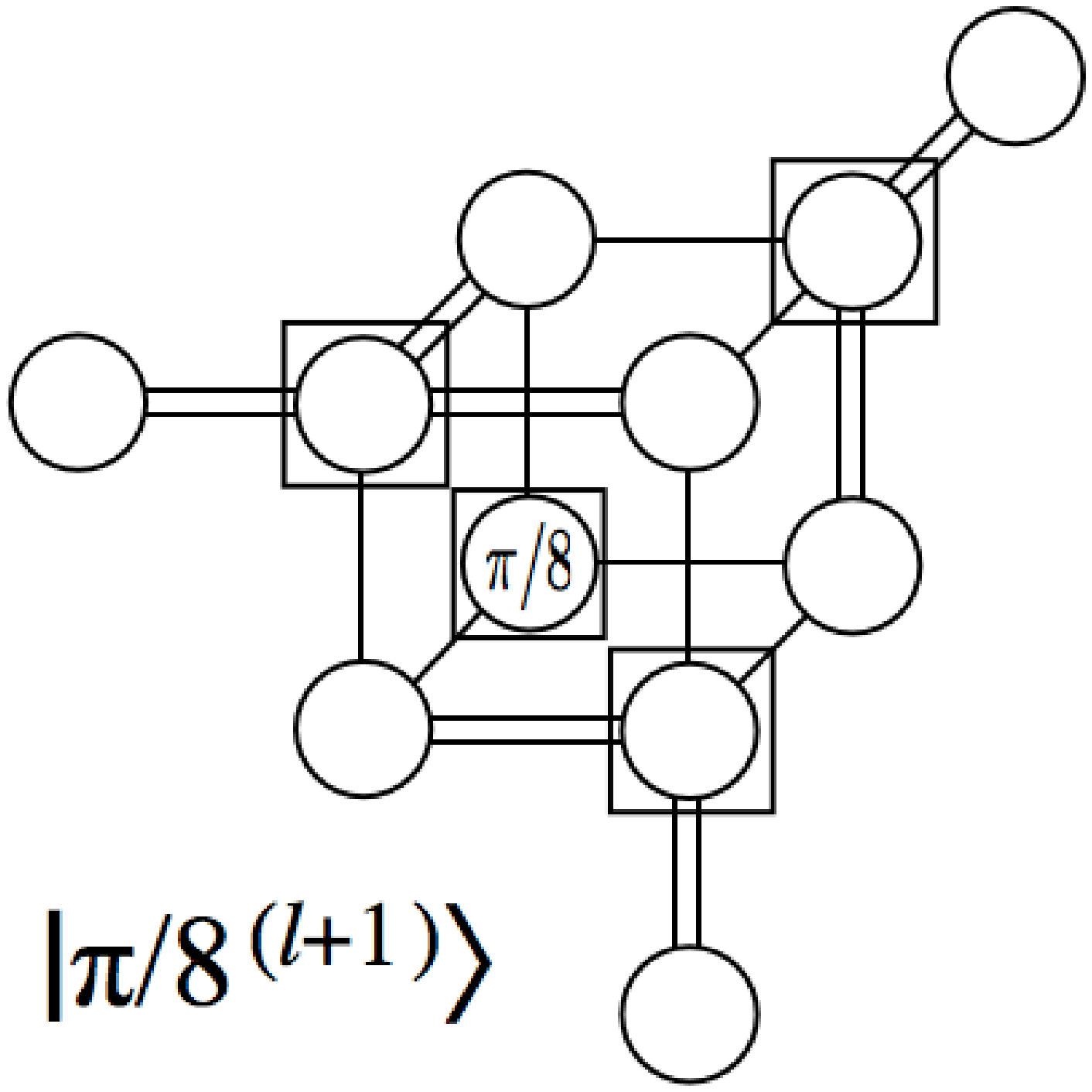}}
\label{pi8reduced}
\end{equation}
where the $\pi/8$ circle indicates
the transfer of $ | \pi/8 ^{(l)}\rangle$ through a $H$ rotation,
similarly to the {\LARGE $ \bullet $} and $ \oplus $ encoding operations.
The logical failure of $ | {5 \pi / 8}^{(l+1)}\rangle $, however,
cannot be detected in the construction of $ | {\pi / 8}^{(l+1)} \rangle $
because it has also the correct syndrome.
Thus, this small mixture of $ | {5 \pi/8}^{(l+1)} \rangle $
is not reduced by the concatenation,
though the constructed $ | {\pi / 8}^{(l+1)} \rangle $
is kept on the code space by verification,
retaining the logical fidelity
as the $ | {\pi / 8}^{(1)} \rangle $.
This slightly noisy $ | {\pi / 8}^{({\bar l})} \rangle $ ($l+1={\bar l}$)
is even useful to obtain the desired high fidelity
$ | {\pi / 8}^{({\bar l})} \rangle $ at the highest level
by using the magic state distillation with Clifford operations
\cite{BK05,Reichardt05}.

\section{Noise threshold}
\label{sec:threshold}

We have described in the previous section
how to construct the verified fundamental clusters in concatenation,
which enables us to implement universal computation fault-tolerantly.
In the following sections we investigate
the performance of this cluster-based architecture,
including a high noise threshold by postselection
and reasonable resources usage for scalability.

The construction of fundamental clusters is performed
via the one-way computation at the lower level.
This provides readily the threshold condition
for the cluster-based architecture:
{\it The error probability for the measurement of each logical qubit,
which composes the verified fundamental clusters,
should be reduced arbitrarily by raising the concatenation level}.
The errors in measuring the logical qubits are twofold;
(i) the errors on the logical qubits themselves,
and (ii) the errors on the Pauli frames,
which are propagated as by-products of one-way computation \cite{OWC}.
The errors of (ii) are thus given by induction
as some multiple of those of (i) in the leading order.
We also note, as discussed in Sec. \ref{sec:simple model},
that the cluster-based architecture exploits a good transversal property
on a suitable code,
which provides, in collaboration with the postselection,
a simple concatenation structure of the logical errors
in the verified fundamental clusters.
Here, we estimate the noise threshold
by considering these features of the cluster-based architecture.
In this calculation we adopt the noise model as follows:

$ \bullet $
A two-qubit gate is followed by $A \otimes B$ errors
with probabilities $p_{AB}$ ($ A, B = I, X, Y, Z $, and $ AB \not= II $).

$ \bullet $
The physical qubits $|0\rangle$ and $|+\rangle$
are prepared as mixed-states with an error probability $p_{p}$:
\begin{eqnarray}
|0\rangle & \rightarrow &
(1-p_{p}) | 0 \rangle \langle 0 | + p_{p} | 1 \rangle \langle 1 | ,
\\
| + \rangle & \rightarrow &
(1-p_{p}) | + \rangle \langle + | + p_{p} | - \rangle \langle - | .
\end{eqnarray}

$ \bullet $
The measurement of a physical qubit in the $A$ ($ X, Y, Z $) basis
is described by positive-operator-valued measure (POVM) elements
$\{ M_{A}^{+} , M_{A}^{-} \}$ with an error probability $p_{M}$:
\begin{eqnarray}
M_{A}^{+} &=& (1-p_{M}) E_{A}^{+} + p_{M} E_{A}^{-} ,
\\
M_{A}^{-} &=& (1-p_{M}) E_{A}^{-} + p_{M} E_{A}^{+} ,
\end{eqnarray}
where $ E_{A}^{\pm} = ( I \pm A )/2 $ are the projectors
to the $\pm 1$  eigenstates of the Pauli operator $A$, respectively.
%

\subsection{Homogeneous errors in verified clusters}

We first consider the errors on the level-0 (physical-level) qubits
encoded in the level-2 fundamental clusters.
Although the correlated errors are introduced
in the encoding process of the level-1 qubits,
they are detected and discarded by postselection sufficiently
through the single and double verifications
in the circuits (\ref{single_circuit}) and (\ref{double_circuit})
for the level-2 cluster construction.
These verification protocols are implemented by the transversal operations.
Thus, it is reasonably expected that the level-0 qubits
encoded in these verified level-1 qubits,
which compose the level-2 fundamental clusters,
have independently and identically distributed (homogeneous)
depolarization errors in the leading order \cite{Eastin07}.
Specifically, the homogeneous error probabilities
$ \epsilon_A $ ($ A = X, Y, Z $) for the level-0 qubits
are determined by those $ p_{AB} $ for the physical two-qubit gates
which are used transversally for the double verification
in the final stage of construction.
This is illustrated in the circuit (\ref{double_circuit}) as
\begin{equation}
\scalebox{.35}{\includegraphics*[0cm,0.5cm][22cm,8cm]{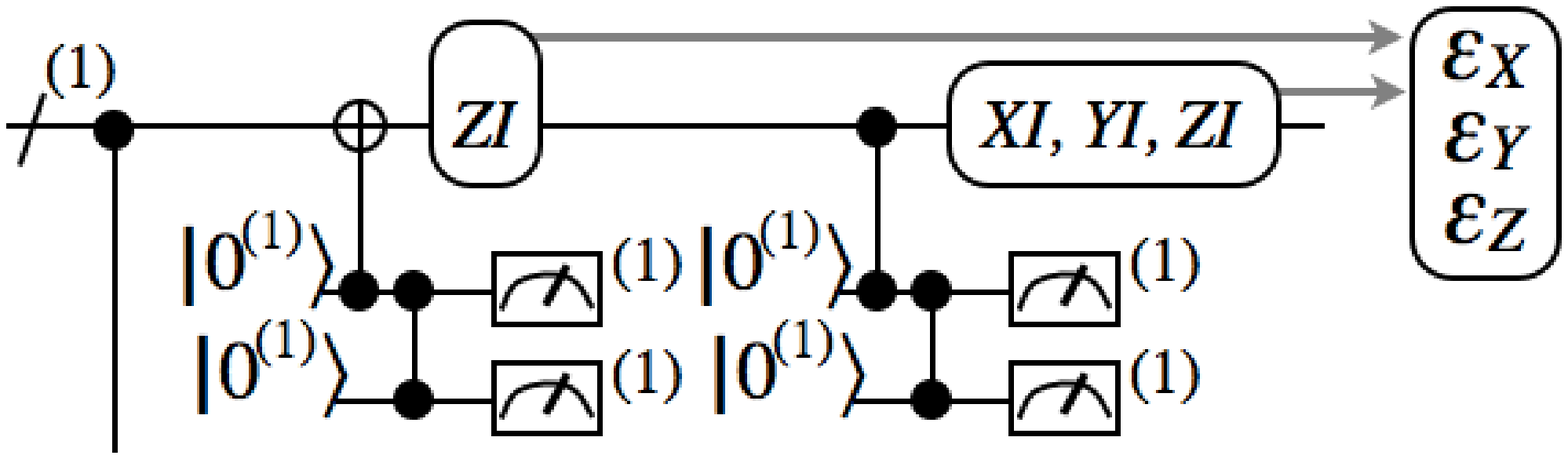}}
\label{homo-error}
\end{equation}
providing the homogeneous errors,
\begin{eqnarray}
\epsilon_X = p_{XI} , \epsilon_Y = p_{YI} , \epsilon_Z &=& 2 p_{ZI} ,
\label{eq-homo-error}
\end{eqnarray}
up to the higher-order contributions.
The errors preceding the double verification,
including the preparation error with $ p_{p} $,
are fully detected and discarded by postselection in the leading order,
as discussed below the circuit (\ref{double_circuit}).

The verified level-2 fundamental clusters are
connected with the transversal bare C$Z$ gates
to construct the level-3 fundamental clusters
as shown in the diagram (\ref{hexa}). 
After the one-way computation with postselection,
the output level-2 qubits are left successfully,
composing the level-3 fundamental clusters.
Here, it should be noted that the output level-2 qubits,
$\circledcirc$'s in the diagram (\ref{hexa}), are never touched directly
in the level-3 cluster construction.
Instead, the entanglement by the verified C$Z$ gates
is transferred via teleportation (one-way computation)
transversally to the output level-2 qubits
to form the verified level-3 fundamental clusters.
Thus, each constituent level-0 qubit in these entangled level-2 qubits
inherits transversally the homogeneous errors $ \epsilon_A $
in Eq. (\ref{eq-homo-error})
after the double verification in the level-2 cluster construction.
The above argument is extended recursively
to the verified level-$l$ fundamental clusters ($l \geq 2$).
As a result, the errors in the verified fundamental clusters
(before the bare C$Z$ connections in the next-level construction)
are reasonably described
in terms of the homogeneous errors $\epsilon _{A}$ on the level-0 qubits.
This fact really simplifies the error structure
in the cluster-based architecture.
Furthermore, the Pauli frame errors are removed
in the leading order for the output qubits through the double verification.
Thus, the cluster-based architecture provides a scalable way
to construct a concatenated code state
whose errors are well approximated by the homogeneous errors,
which was assumed in Ref. \cite{Eastin07}.

\subsection{Noise threshold calculation}

We next consider the errors for the measurement of the logical qubits
in the one-way computation to construct the verified fundamental clusters.
The level-$ l $ clusters with the homogeneous errors $\epsilon _{A}$
on their constituent level-0 qubits
are used for the level-$ (l+1) $ cluster construction.
As seen in the previous section, e.g.,
the diagram (\ref{single-verification-full}),
some pairs of level-$ (l-1) $ qubits in these level-$l$ clusters
are connected by the bare C$Z$ gates.
As a result, extra errors are added transversally
to the constituent level-0 qubits through the bare C$Z$ connection,
as shown in the following diagram:
\begin{equation}
\scalebox{.35}{\includegraphics*[0cm,0.5cm][18cm,9cm]{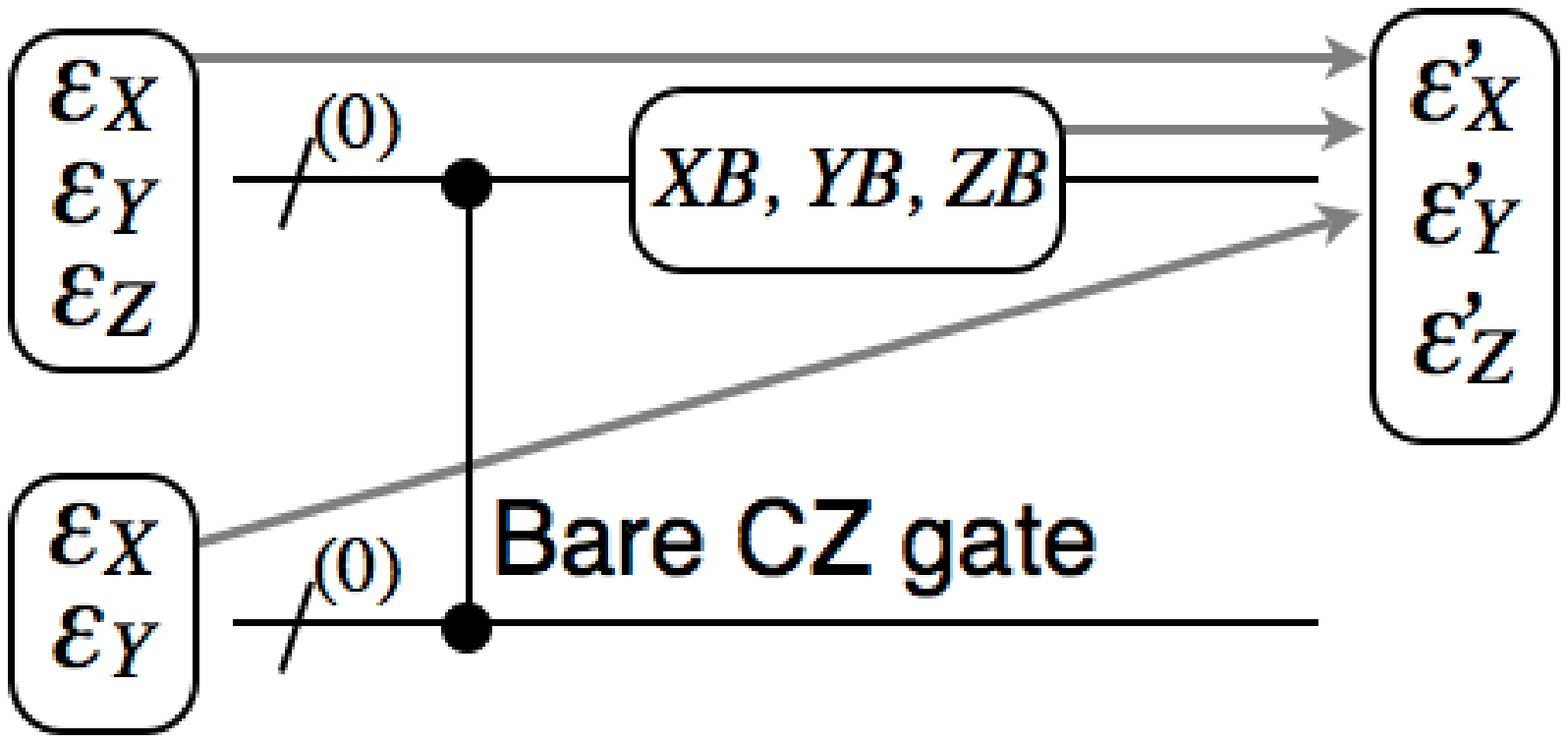}}
\label{bare_C_Z_error}
\end{equation}
Then, the homogeneous errors after the bare C$Z$ connection
are given in the leading order as
\begin{eqnarray}
\epsilon^\prime_X &=& \epsilon_X + \sum_{B= I, X, Y, Z} p_{XB} ,
\label{hoge-bare-errorX}
\\
\epsilon^\prime_Y &=& \epsilon_Y + \sum_{B= I, X, Y, Z} p_{YB} ,
\label{hoge-bare-errorY}
\\
\epsilon^\prime_Z &=& \epsilon_Z + \epsilon_X + \epsilon_Y
+ \sum_{B= I, X, Y, Z} p_{ZB} .
\label{hoge-bare-errorZ}
\end{eqnarray}

Now we are ready to calculate the error probability for the measurement
of the bare-connected level-$l$ qubit
which is implemented in concatenation by the transversal measurements
of the constituent lower-level qubits.
Consider first the level-1 qubits
composing the level-2 fundamental clusters,
which are measured in the level-1 one-way computation
for the level-3 cluster construction.
Note here that the level-0 qubits (constituents of the level-1 qubits)
are not assigned the Pauli frames
in the circuit-model computation at the physical level
to construct the level-2 fundamental clusters.
(Even if the cluster-model computation is adopted at the physical level,
the Pauli frame error can be neglected in a good approximation,
which is left only as the second-order error contribution
after the double verification.)
Thus, the measurement of the level-1 qubit is affected
by the errors $ \epsilon^\prime_{A} $ on the level-0 qubits
and the physical measurement error $ p_{M} $.
The logical error probability for the $X$ measurement
of the bare-connected level-1 qubit is then calculated
in the leading order on the Steane seven-qubit code with distance 3 as
\begin{eqnarray}
p_{q}^{(1)} \simeq {}_7 {\rm C}_2
( \epsilon'_{Z} + \epsilon'_{Y} + p_{M} )^2
\equiv {}_7 {\rm C}_2 ( p_{q}^{(0)} )^2 ,
\label{level-1pq}
\end{eqnarray}
where $p_{q}^{(0)}$ is defined as the error probability
for the $X$ measurement of the bare-connected level-0 qubit.
It is apparent here that
by choosing properly the physical basis
the errors for the $Z$ and $Y$ measurements
are arranged to be smaller than $p_{q}^{(0)}$ for the $X$ measurement,
i.e., $ \epsilon'_{Z} \geq \epsilon'_{Y} \geq \epsilon'_{X} $.

The outcomes of the measurements of the level-1 qubits
are propagated to the neighboring qubits by updating the Pauli frames
according to the rule of one-way computation \cite{OWC}.
Then, the errors on the measurement outcomes
with the probability $p_{q}^{(1)}$ are accumulated during the computation.
The blocks of seven output level-1 qubits (level-2 qubits)
to form the level-3 fundamental clusters
are, however, doubly verified in the final stage of one-way computation.
Thus, the propagation of the preceding measurement errors
as the Pauli frame error is prohibited by postselection in the leading order
for these output level-1 qubits,
as discussed in the circuit (\ref{double_circuit}):
\begin{equation}
p_{\rm Pauli}^{(1)} \sim (p_{q}^{(1)})^{2} .
\label{p_Pauli}
\end{equation}

Subsequently, the level-2 one-way computation is performed
by using the level-3 fundamental clusters
to construct the level-4 fundamental clusters,
where the constituent level-2 qubits are measured.
Some of the level-2 qubits are connected
with the transversal bare C$Z$ gates for the first time in this computation.
The measurement of the (bare-connected) level-2 qubit is executed
by measuring the (bare-connected) level-1 qubits transversally.
The seven level-1 measurement outcomes
together with the seven level-1 Pauli frames
determine the level-2 measurement outcome.
Then, by considering Eq. (\ref{p_Pauli})
the error probability for measuring the level-2 qubit
after the bare C$Z$ connection is given in the leading order as
\begin{eqnarray}
p_{q}^{(2)} \simeq {}_7 {\rm C}_2 ( p_{q}^{(1)} + p_{\rm Pauli}^{(1)} )^2
\simeq {}_7 {\rm C}_2 ( p_{q}^{(1)} )^2 .
\label{level-2pq}
\end{eqnarray}
As for the logical error left on the Pauli frame of each output qubit
after the cluster construction,
similarly to Eq. (\ref{p_Pauli}),
it is reduced by the double verification as
\begin{eqnarray}
p_{\textrm{Pauli}}^{(l-1)} \sim (p_{q}^{(l-1)})^2 ( l \geq 2 ) .
\end{eqnarray}
Thus, the error probability $ p_q^{(l)} $
for measuring the level-$ l $ qubit is calculated in concatenation as
\begin{eqnarray}
p_q^{(l)} 
\simeq {}_7 {\rm C}_2 ( p_q^{(l-1)} )^2 
\simeq ( {}_7 {\rm C}_2 p_q^{(0)} )^{2^l} / {}_7 {\rm C}_2 .
\label{eqn:pq}
\end{eqnarray}
The threshold condition is then given from Eq. (\ref{eqn:pq}) as
\begin{equation}
p_q^{(0)} = D p_{g} < 1/{}_7 {\rm C}_2 ,
\end{equation}
and the noise threshold is estimated as
\begin{equation}
p_{\rm th} \simeq ( {}_7 {\rm C}_2 D )^{-1} ,
\label{eqn:pth}
\end{equation}
where $ p_{g} $ represents the mean error probability
for physical operations ($ D \sim 1 $).
Typically with $ p_{AB} = (1/15) p_{g} $
for $ \epsilon^\prime_{A} $ and $ p_M = (4/15) p_{g} $ \cite{Knill05b},
where $ D = 17/15 $, the noise threshold is estimated
approximately as $ p_{\rm th} \simeq 0.04 $.

\subsection{Numerical simulation}

We have made numerical calculations to confirm the above estimation
of the error probability $ p_q^{(l)} $ for measuring the logical qubit
and the noise threshold $ p_{\rm th} $ for computation
by simulating the construction of fundamental clusters.

\begin{figure}
\scalebox{1}{\includegraphics*[1cm,1.5cm][10cm,6.5cm]{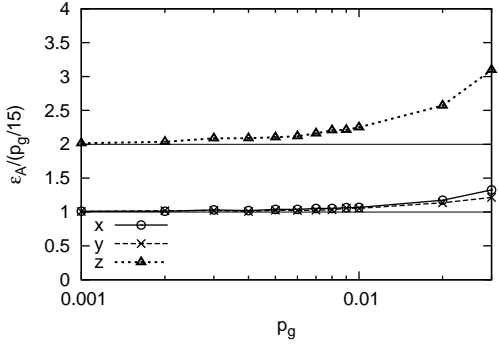}}
\caption{The error probabilities
$\epsilon_{A}/(p_{g}/15)$ ($A=X,Y,Z$) for each level-0 qubit
are plotted as functions of the physical error probability $p_{g}$
together with their leading values
$ \epsilon_{X} (p_g /15) = \epsilon_{Y}/(p_g /15) = 1 $
and $ \epsilon _{Z}/(p_g /15) = 2 $.}
\label{epsilon}
\end{figure}
\begin{figure}
\scalebox{.9}{\includegraphics*[1.5cm,1.5cm][12cm,8.8cm]{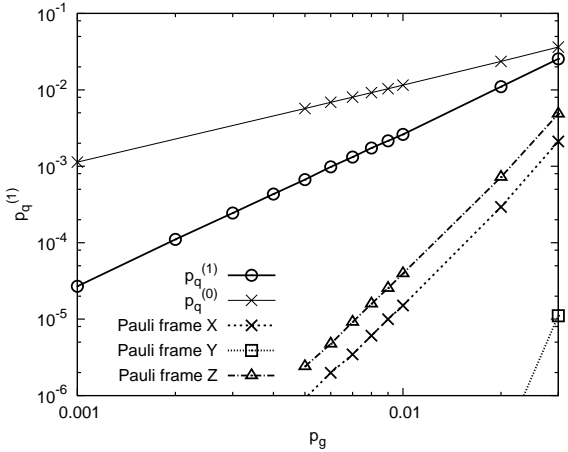}}
\caption{The error probability $p^{(1)}_{q}$
for measuring the level-1 qubit after the bare C$Z$ connection
is plotted as a function of the physical error probability $p_g$.
The error probabilities $p_{\rm Pauli}^{(1)}$ 
for the Pauli frames ($X, Y, Z$) of the level-1 qubit
are also plotted as functions of $p_g$ in comparison with $p^{(1)}_{q}$.
The upper-most line indicates $p^{(0)}_{q} $
in comparison to infer the threshold.}
\label{simulation}
\end{figure}

First, we have constructed the level-2 fundamental clusters
according to the diagrams (\ref{hexa-reduced}) and (\ref{qubit-reduced})
by implementing the C$Z$ operations
with single and double verifications for the level-1 encoded qubits
in the circuits (\ref{single_circuit}) and (\ref{double_circuit})
with bare C$Z$ gates (transversal operation of physical C$Z$ gates).
Then, we have checked the error probabilities $ \epsilon_{A} $
($A=X,Y,Z$) for each level-0 qubit
which is contained in the output level-1 qubits
as the verified level-2 fundamental clusters.
In Fig. \ref{epsilon} $ \epsilon_{A} / (p_{g}/15) $
are plotted as functions of the physical error probability $p_{g}$,
where $ p_{AB} = p_{g}/15 $, $ p_M = (4/15) p_{g} $
and $ p_{p} = (4/15) p_{g} $ \cite{Knill05b} are specifically adopted.
In the case of $p_{g} < 1\%$
they are in good agreement with the leading values
$ \epsilon_{X}/(p_{g}/15) = \epsilon_{Y}/(p_{g}/15) = 1 $
and $ \epsilon_{Z}/(p_{g}/15) = 2 $ in Eq. (\ref{eq-homo-error}).
On the other hand, in the case of $p_{g}>1\%$
$\epsilon_{A}/(p_{g}/15)$ become larger
due to the higher-order contributions,
which are thus significant for $p_{q}^{(1)}$.
It has been also checked for $p_{g} \leq 3 \% $
that these errors are almost independent among the level-0 qubits;
the correlated errors are one order of magnitude smaller
than the independent ones
even when the higher-order contributions are significant
for $ \epsilon_{A} $.
We have then evaluated the error probability $p_{q}^{(1)}$
for measuring the output level-1 qubit
(component of the level-2 fundamental cluster)
after operating the bare C$Z$ gate on it.
It is plotted in Fig. \ref{simulation} as a function of $p_{g}$.

Next, we have constructed the level-3 fundamental clusters
by simulating the one-way computation
for the level-1 qubits (level-2 cluster states)
in the diagrams such as (\ref{hexa})
or their full three-dimensional versions.
Then, we have calculated the error probabilities $p_{\rm Pauli}^{(1)}$
for the Pauli frames ($X, Y, Z$) of the level-1 qubit
which is contained in the output level-2 qubit
(component of the level-3 fundamental cluster).
They are plotted in Fig. \ref{simulation} as functions of $p_{g}$
in comparison with the error probability $p_{q}^{(1)}$
for measuring the level-1 qubit.
This result really confirms that $p_{\rm Pauli}^{(1)}$
is suppressed substantially by the double verification,
to be of the second order of $p_{q}^{(1)}$, as shown in Eq. (\ref{p_Pauli}).

\begin{figure}
\scalebox{1.0}{\includegraphics*[1.5cm,1.5cm][12.5cm,7.5cm]
{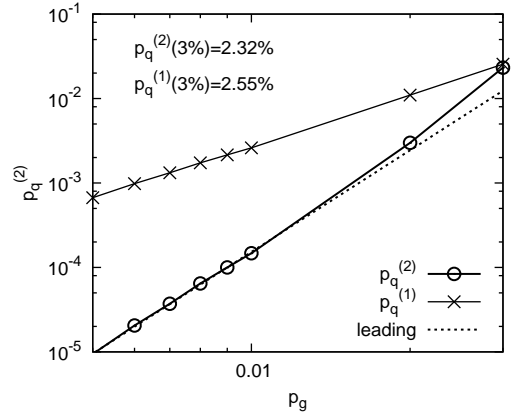}}
\caption{The error probability $p^{(2)}_{q}$
for measuring the level-2 qubit after the bare C$Z$ connection
is plotted as a function of the physical error probability $p_g$,
together with the leading term $ {}_7 {\rm C}_2 ( p_{q}^{(1)} )^2 $
(dotted line).
The upper-most line indicates $ p^{(1)}_{q} $
in comparison to infer the threshold.}
\label{simulation2}
\end{figure}
By using these values of $p_{q}^{(1)}$ and $p_{\rm Pauli}^{(1)}$
for the level-1 qubit, we have calculated
the error probability $ p^{(2)}_{q} $ for measuring the output level-2 qubit
(component of the level-3 fundamental cluster)
after the bare C$Z$ connection.
It is plotted as a function of $p_{g}$ in Fig. \ref{simulation2}
together with the leading term $ {}_7 {\rm C}_2 ( p_{q}^{(1)} )^2 $
(dotted line) as given in Eq. (\ref{level-2pq}).
(The error effect for $ p^{(2)}_{q} $ due to the bare C$Z$ connection
is already taken into account transversally
as a contribution in $p_{q}^{(1)}$.)
Here, it is found that for $ p_{g} > 1\% $ near the threshold
the level-2 qubit error $p^{(2)}_{q}$
becomes significantly higher than its leading value (dotted line)
due to the higher-order contributions including the Pauli frame error.
The logical error probability, however, decreases
through concatenation as $p^{(2)}_{q} < p^{(1)}_{q} < p^{(0)}_{q} $
for $ p_{g} \leq 3 \% $.
This certainly indicates that the noise threshold $ p_{\rm th} $
is about $ 3 \% $,
which is in reasonable agreement with the leading-order estimate
in Eq. (\ref{eqn:pth}).
The noise threshold $ p_{\rm th} \sim 3\% $ of the present architecture
is considerably higher than those
of the usual circuit-based architectures with the Steane seven-qubit code.
It is also comparable to those of the two $C_{4}/C_{6}$ architectures,
error-correcting and postselecting ones \cite{Knill05a}.

\section{Resources usage}
\label{sec:resources}

The physical resources (qubits and gates) are calculated
by counting the numbers of hexa-clusters, ancilla code states
and bare C$Z$ gates which are used in the diagrams
for the construction of fundamental clusters.
In this calculation we present recursion relations
of the resources $R^{(l)}_{\alpha}$
required for the components $ \alpha = S, D, h, 0, + $
corresponding to the single verification, double verification,
hexa-cluster $|h\rangle$,
ancilla qubits $|0\rangle$ and $|+\rangle$, respectively.

The single verification in the diagram (\ref{single-cluster})
or its full version (\ref{single-verification-full})
uses 1$\times$7 $|h^{(l)}\rangle$'s, 2 $|+^{(l)}\rangle$'s
and 2 level-$l$ transversal bare C$Z$ gates, that is
\begin{eqnarray}
R_S^{(l)} &=& 1 \times 7 R_h^{(l)} + 2 ( R_+^{(l)} + R_b^{(l)} )
( l \geq 2 ) ,
\label{eqn:RS}
\end{eqnarray}
where
\begin{eqnarray}
R_{b}^{(l)} = 7^{l}
\end{eqnarray}
indicates the resources for a level-$l$ transversal bare C$Z$ gate
(the number of physical C$Z$ gates).
Similarly, the resources $R^{(l)}_{D}$ for the double verification,
which uses $3 \times 7$ $|h^{(l)}\rangle$'s,
8 $|+^{(l)}\rangle$'s and and $(8+2)$ level-$l$ bare C$Z$ gates
in the diagram (\ref{double-cluster}), are given as
\begin{equation}
R_D^{(l)} = 3 \times 7 R_h^{(l)} + 8 ( R_+^{(l)} + R_b^{(l)} )
+ 2 R_b^{(l)} ( l \geq 2 ) .
\label{eqn:RD}
\end{equation}
Furthermore, the resources used
to construct the level-$(l+1)$ fundamental clusters
$|h^{(l+1)}\rangle$, $|0^{(l+1)}\rangle$ and $|+^{(l+1)}\rangle$
are counted from the reduced diagrams (\ref{hexa-reduced})
and (\ref{qubit-reduced}) as
\begin{eqnarray}
R_\alpha^{(l+1)}
&=& \sum_{\beta = S,D,0,b}
\frac{n^\beta_\alpha R_\beta^{(l)}}{p_\alpha^{(l+1)}}
( \alpha = h, 0, +; l \geq 1 ) ,
\label{eqn:Ralpha}
\end{eqnarray}
with the numbers of the respective level-$l$ components
\begin{eqnarray}
(n^{S}_{h},n^{D}_{h},n^{0}_{h},n^{b}_{h}) &=& ( 2, 3, 6, 10 ) ,
\\
(n^{S}_{0},n^{D}_{0},n^{0}_{0},n^{b}_{0}) &=& ( 6, 7, 11, 26 ) ,
\\
(n^{S}_{+},n^{D}_{+},n^{0}_{+},n^{b}_{+}) &=& ( 5, 7, 10, 24 ) ,
\end{eqnarray}
and the success probabilities $ p_\alpha^{(l+1)} $
for the clusters $ |\alpha^{(l+1)}\rangle $
to pass the verification process with postselection.
Here, the bare C$Z$ gates are used in the processes,
(i) the $n^{0}_{\alpha}$ encodings with $|0^{(l)}\rangle$ ($\oplus$),
and (ii) the $[2(n^{S}_{\alpha}+n^{D}_{\alpha}) - n^{0}_{\alpha}]$
connections between the outputs after the verifications
and the inputs to the subsequent verifications,
where $ n^{0}_{\alpha} $ is subtracted for the final outputs
({\large $ \circledcirc $}).
Thus, the number of the level-$l$ bare C$Z$ gates is given
by $ n^{b}_{\alpha} = 2(n^{S}_{\alpha}+n^{D}_{\alpha}) $, i.e.,
$n^{b}_{h}=10$, $n^{b}_{0}=26$ and $n^{b}_{+}=24$.
The bare C$Z$ gates are also used in the verification diagrams,
which are properly counted in $R_{S}^{(l)}$ and $R_{D}^{(l)}$.
The level-1 resources are given
in the circuits (\ref{single_circuit}), (\ref{double_circuit})
and (\ref{qubit_level1}) as
\begin{eqnarray}
R_S^{(1)} &=& 3 R_{b}^{(1)} + 2 R_0^{(1)} ,
\\
R_D^{(1)} &=& 9 R_{b}^{(1)} + 8 R_0^{(1)} ,
\\
R_0^{(1)} &=&  R_{+}^{(1)}= 69 / p_0^{(1)} .
\end{eqnarray}
Here, $ R_{0,+}^{(1)} $ is counted as follows.
The Steane seven-qubit code state is encoded into 7 physical qubits
by using 9 CNOT gates \cite{Stea98}.
This code state is preliminarily verified
through 3 stabilizer measurements,
each of which consumes 1 ancilla qubit and 4 CNOT gates.
At this stage $ 7 + 9 + 3 \times ( 1 + 4 ) = 31 $ resources are used
for each preliminarily verified code state.
Then, the code sate is secondly verified according
to the circuit (\ref{qubit_level1}),
where 2 preliminarily verified code states
and 7 (transversal) CNOT gates are used.
Thus, the number of resources used to prepare the level-1 code state
amounts to $ R_{0,+}^{(1)} = ( 2 \times 31 + 7 ) / p_0^{(1)}
= 69 / p_0^{(1)} $ including the success probability
$  p_0^{(1)} =  p_+^{(1)} $.

\begin{figure}
\scalebox{.8}{\includegraphics*[1.5cm,1.5cm][10cm,8.5cm]
{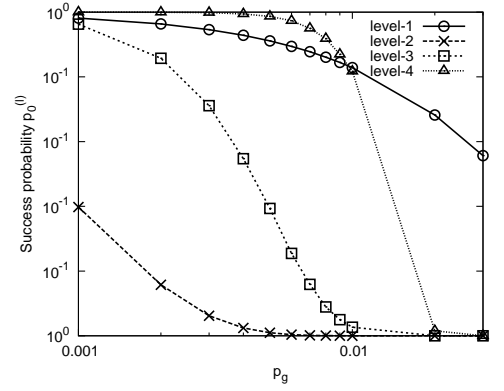}}
\caption{The success probabilities $p_{0}^{(l)}$
are plotted as functions of the physical error probability $p_{g}$
for the levels $l=1,2,3,4$.}
\label{success_probability}
\end{figure}

The success probabilities $ p_\alpha^{(l)} $ have been evaluated
in the numerical simulation for the cluster construction.
In Fig. \ref{success_probability}
we plot especially $p_{0}^{(l)}$ ($\leq p_{+}^{(l)}<p_{h}^{(l)}$)
as functions of the physical error probability $p_{g}$
for the levels $l=1,2,3,4$.
The level-1 $ p_{\alpha}^{(1)} $ appears to be rather high
since the physical-level computation is implemented in the circuits
with less operations.
Then, the level-2 $ p_{\alpha}^{(2)} $ decreases substantially
due to the low fidelity of the level-2 fundamental clusters
for the level-3 cluster construction.
However, the success probabilities $ p_{\alpha}^{(l)}$
almost approach unity at the levels 4 and higher
as the error probability $ p_q^{(l)} $ for the logical qubit
is reduced rapidly for $ p_{g} < 1 \% $ below the threshold.

\begin{figure}
\centering
\scalebox{1.1}{\includegraphics*[1cm,1.75cm][15cm,7cm]{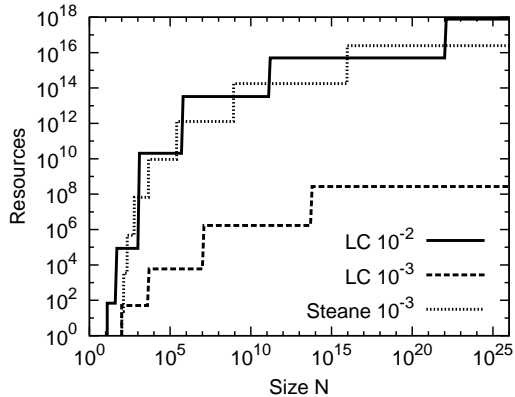}}
\caption{Resources for the present architecture of verified logical clusters
(LC) with $ p_{g} = 10^{-2} $ and $ 10^{-3} $,
which are compared with those for the Steane's QEC scheme
with $ p_{g} = 10^{-3} $.
}
\label{resources}
\end{figure}

The resources are evaluated by using the above recursion relations
with the success probabilities $ p_\alpha^{(l)} $ simulated numerically,
depending on the computation size $ N $,
where the highest level is given
as $ {\bar l} \sim \log_2 ( \log_{10} N ) $
to achieve the accuracy $ 0.1/N $.
The results of $ R_0^{({\bar l})} $ ($ > R_{h,+}^{({\bar l})} $)
are shown in Fig. \ref{resources}
for the present architecture of verified logical clusters (LC)
with $ p_{g} = 10^{-2} $ and $ 10^{-3} $,
which are compared with
the resources for the circuit-based Steane's QEC scheme
with $ p_{g} = 10^{-3} $ \cite{Stea03}.
Each step in these graphs indicates
the rise of the highest level ${\bar l}$ by one.
We find that the present architecture really consumes much less resources
than the Steane's QEC scheme for $ p_{g} \leq 10^{-3} $
(checked numerically also for $p_{g} =10^{-4}$).
This indicates that the overhead costs paid
for the verification process with postselection in the cluster construction
are worth enough to save the total resources usage
by reducing rapidly the logical error probability.
Thus, the present cluster-based architecture
is quite efficient with respect to both noise threshold and resources usage,
compared with the usual circuit-based QEC schemes
with the Steane seven-qubit code.

We also compare the present architecture
with the postselecting and error-correcting $C_{4}/C_{6}$ architectures
\cite{Knill05a}.
The postselecting $C_{4}/C_{6}$ architecture
makes use of the usual circuit-based non-determinism
for fault-tolerant gate operation,
which is different from the error-precorrection
in the cluster-based architecture.
It is thus lacking in scalability by itself,
requiring the construction of a large QEC code state at a certain level
with the decoding of the lower-level error-detection code,
in order to implement the standard fault-tolerant computation
at the higher levels.
The resources usage of the postselecting $C_{4}/C_{6}$ architecture
amounts to be quite large as $ \sim 10^{100} - 10^{1000} $
for the overhead cost of the large QEC code state.
On the other hand, the noise threshold and resources usage
for the error-correcting $C_{4}/C_{6}$ architecture
with the Fibonacci scheme are both comparable
to those for the present cluster-based architecture
with the Steane seven-qubit code.

\section{Miscellaneous}
\label{sec:miscellaneous}

We further discuss some issues
concerning the performance of the cluster-based architecture.

\subsection{Memory error effect}

The memory errors may be significant
in the cluster-based architecture without recovery operation.
The qubits to form the clusters
are not touched directly (but via one-bit teleportation)
through the concatenated constructions after the level-1 verification.
Then, the memory errors accumulate
until they are measured in the upper-level construction.
The memory errors are added as $ p_q^{(0)} + {\bar l}( n \tau_m p_{g} ) $,
where $ \tau_m p_{g} $ denotes the probability of memory error
with the effective waiting time $ \tau_m $ for one measurement,
and $ n $ is the number of waiting time steps
at each concatenation level (e.g., $ n = 12 $ for the hexa-cluster).
The noise threshold is thus estimated roughly as
\begin{equation} 
p_{\rm th}
\sim [ {}_{7}{\rm C}_2 \{ 1 + \log_2 ( \log_{10} N ) n \tau_m \} ]^{-1} ,
\end{equation}
depending on the computation size $ N $
with the highest level $ {\bar l} \sim \log_2 ( \log_{10} N ) $.
For example, $ p_{\rm th} \sim 1 \% $
for $ N \sim 10^{20} $ and $ \tau_m = 0.1 $ ($ n \sim 10 $),
which will be tolerable for practical computations.
In order to overcome essentially the memory error accumulation,
the fundamental clusters as two-colorable graph states
may be refreshed at each level
by using a purification protocol \cite{DAB03ADB05,Goyal06}.

\subsection{One-way computation at the physical level}
We may use the one-way computation even at the physical level,
instead of the circuit computation,
for the construction of level-2 fundamental clusters.
The level-1 qubits are encoded through the verification
by the cluster versions of the circuits in (\ref{qubit_level1}).
Then, the level-2 hexa cluster is constructed
through the single and double verifications
as given in the reduced diagram in (\ref{hexa-reduced})
by combining the physical qubits and level-1 code states
with the transversal bare C$Z$ gates:
\begin{equation}
\scalebox{.35}{\includegraphics*[0cm,0cm][15cm,14.5cm]{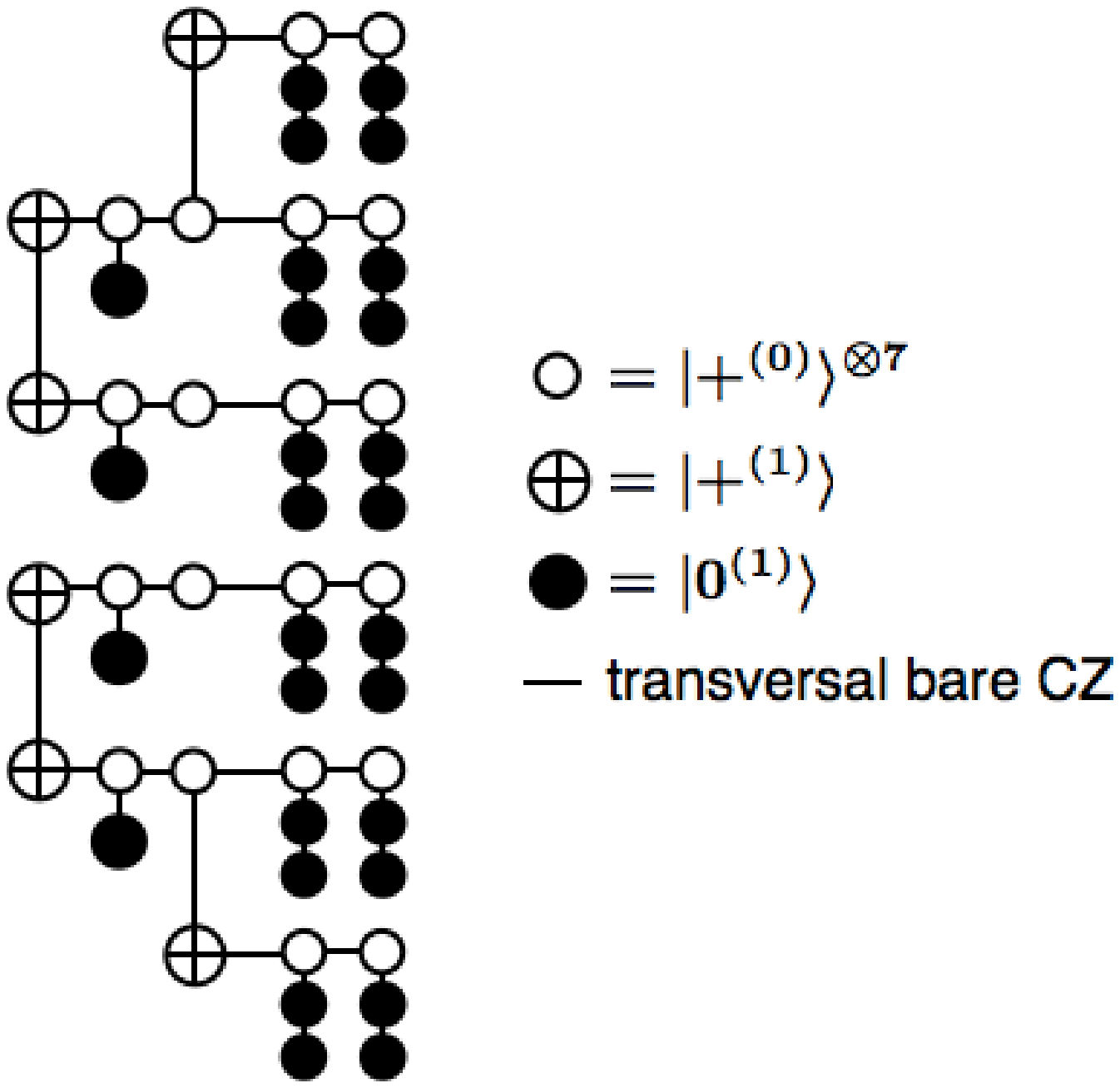}}
\label{HexaCZ}
\end{equation}
The level-2 code states are constructed similarly
according to the reduced diagrams in (\ref{qubit-reduced}).
The homogeneous errors for the resultant level-1 qubits
(components of the level-2 clusters)
are estimated in the first order
by inspecting the double verification process in the final stage,
where extra C$Z$ gates are required for the CNOT gate operations
inducing additional errors:
$ \epsilon_{X} = p_{XI} $, $ \epsilon_{Y} = p_{YI} $,
$ \epsilon_{Z} = p_{p} + p_{XZ} + p_{IZ} + p_{ZY}
+ p_{YY} + p_{ZI} + p_{ZI} $.
The noise threshold is slightly lowered
as $ p_{\rm th} \simeq 0.03 $ with $ D = 5/3 $ in Eq. (\ref{eqn:pth}).

\subsection{Application of other QEC codes}

So far we have considered only the Steane seven-qubit code
in the present architecture.
Here, we briefly discuss application of some other QEC codes, say code $C$.
If the code $C$ is a self-dual CSS code
or a CSS code which has high symmetry
such as the Bacon-Shor subsystem code,
the cluster-based architecture can be applied straightforwardly
by taking the hexa-cluster
and the graph state equivalents of the code states of $C$
as the fundamental clusters.
The behavior of logical errors is, however, somewhat different,
depending on the distance of $C$ as seen in the following two examples.

We first consider the four-qubit error detection code $ C_{4} $.
The Fibonacci scheme can be used for the $ C_{4} $ code
to generate deterministically the logical measurement outcomes
from the physical ones in one-way computation.
Then, the cluster-based concatenation
can be carried out with the error detection code $C_{4}$
almost in the same way as with the Steane's seven-qubit code.
In this case, we may reduce the resources
to prepare the level-2 fundamental clusters
with high success probability,
since the number of error locations is smaller
than that for the Steane seven-qubit code \cite{Knill05a,Aliferis}.
As a trade-off the error probability for the Pauli frame
becomes $p_{\rm Pauli}^{(1)} \sim p_{g}^3$,
while the error probability for measuring the level-1 qubit
is $p_{q}^{(1)} \sim p_{g}^2$.
Thus, the Pauli frame provides a more significant error contribution
near the threshold than the case of the Steane seven-qubit code
with $p_{\rm Pauli}^{(1)} \sim p_{g}^4$.

We next consider the Golay code,
which is a 23-qubit self-dual CSS code with distance 7.
In this case, although we have to pay much more resources
at the lowest level, the logical errors are reduced substantially
as $p_{q}^{(1)} \sim p_{g}^4$ and $p^{(1)}_{\rm Pauli}\sim p_{g}^8$
\cite{Stea03,Cross07}.
Thus, it will be possibility to improve the noise threshold
of the cluster-based architecture by using the Golay code.

We further mention that even with the Steane seven-qubit code
the present architecture has a room to improve its performance.
The optimal decoding (adaptive concatenation) technique
\cite{Poulin06Fern08},
which boosts the correctable error of the Steane seven-qubit code
up to $\sim 11 \%$, is readily available to improve the noise threshold
by generating efficiently the logical measurement outcomes
in one-way computation.

\section{Conclusion}
\label{sec:conclusion}

We have investigated an efficient architecture
for fault-tolerant quantum computation,
which is based on the cluster model of encoded qubits.
Some relevant logical cluster states, fundamental clusters,
are constructed through verification without recovery operation
in concatenation, which provides the error-precorrection
of gate operations for the one-way computation at the higher level.
A suitable code such as the Steane seven-qubit code
is adopted for transversal operations.
This construction of fundamental clusters
provides a simple transversal structure of logical errors in concatenation,
and achieves a high noise threshold
by using appropriate verification protocols,
namely the single and double verifications.
Since the postselection is localized within each fundamental cluster
with the help of deterministic bare C$Z$ gates without verification,
divergence of resources is restrained, which reconciles
postselection with scalability.
Detailed numerical simulations have really confirmed
these desired features of the cluster-based architecture.
Specifically, the noise threshold is estimated to be about $ 3 \% $,
and the resources usage is much less
than those of the usual circuit-based QEC schemes
with the Steane seven-qubit code.
This performance is comparable
to that of the error-correcting $C_{4}/C_{6}$ architecture
with the Fibonacci scheme.
Some means may hopefully be applied
for the cluster-based architecture to improve its performance,
including the error-detecting $C_{4}$ code with the Fibonacci scheme,
other self-dual CSS codes such as the Golay code,
which are more robust for logical encoding
than the Steane seven-qubit code,
and the adoptive concatenation or optimal decoding.

\begin{acknowledgments}
This work was supported by the JSPS Grant No. 20.2157.
\end{acknowledgments}

\appendix

\section{Diagrams for cluster construction}
\label{sec:code state}

We can produce systematically the diagrams for cluster construction
from the reduced ones (\ref{hexa-reduced}) and (\ref{qubit-reduced}),
according to the following rules:
(i) Replace the single edge with the single verification
(\ref{single-cluster}).
(ii) Replace the double edge with the double verification
(\ref{double-cluster})
so that the double verifications are always placed
at the right side (namely later in time) of the single verifications.
(iii) Put the $\oplus$ encodings on the input qubits at the leftmost
(initially in time).
(iv) Apply the bare C$Z$ gate (wavy line)
to connect the output qubit of the preceding verification
to the input qubit of the following verification.
In the case that the double verification
is followed by the other double verification,
we cut off the leftmost qubit of the following verification by measurement
before connecting these double verifications,
in order to remove the redundant $H$ rotation.
This prescription is illustrated in the following diagram:
\begin{equation}
\scalebox{.28}{\includegraphics*[0cm,0.5cm][29cm,11cm]{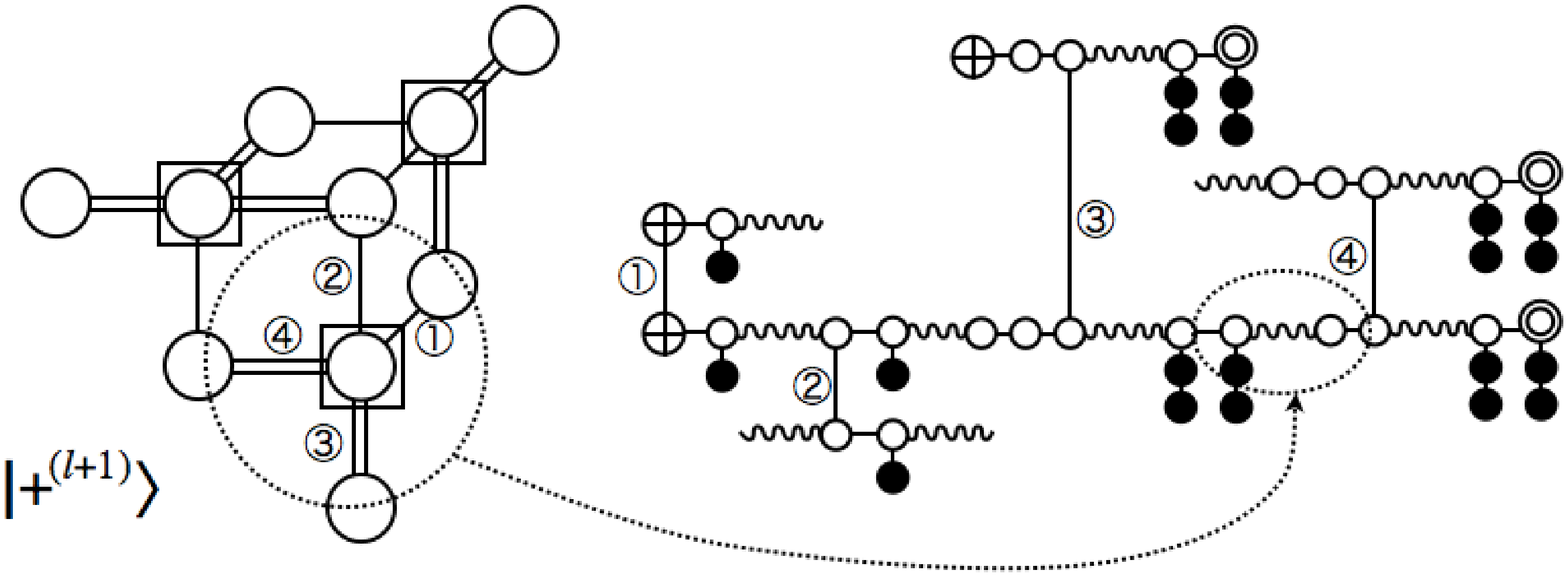}}
\label{reduced-rules}
\end{equation}
The cluster diagram (\ref{hexa}) for $|h^{(l+1)}\rangle$
is generated according to these rules (i)--(iv).
The cluster states for $|+^{(l+1)}\rangle$ and $|0^{(l+1)}\rangle$
are constructed similarly in the following diagrams,
where the pairs of the same characters such as (a)-(a)
are actually connected by the bare C$Z$ gates:
\begin{equation}
\scalebox{.28}{\includegraphics*[0cm,0.5cm][29cm,22cm]{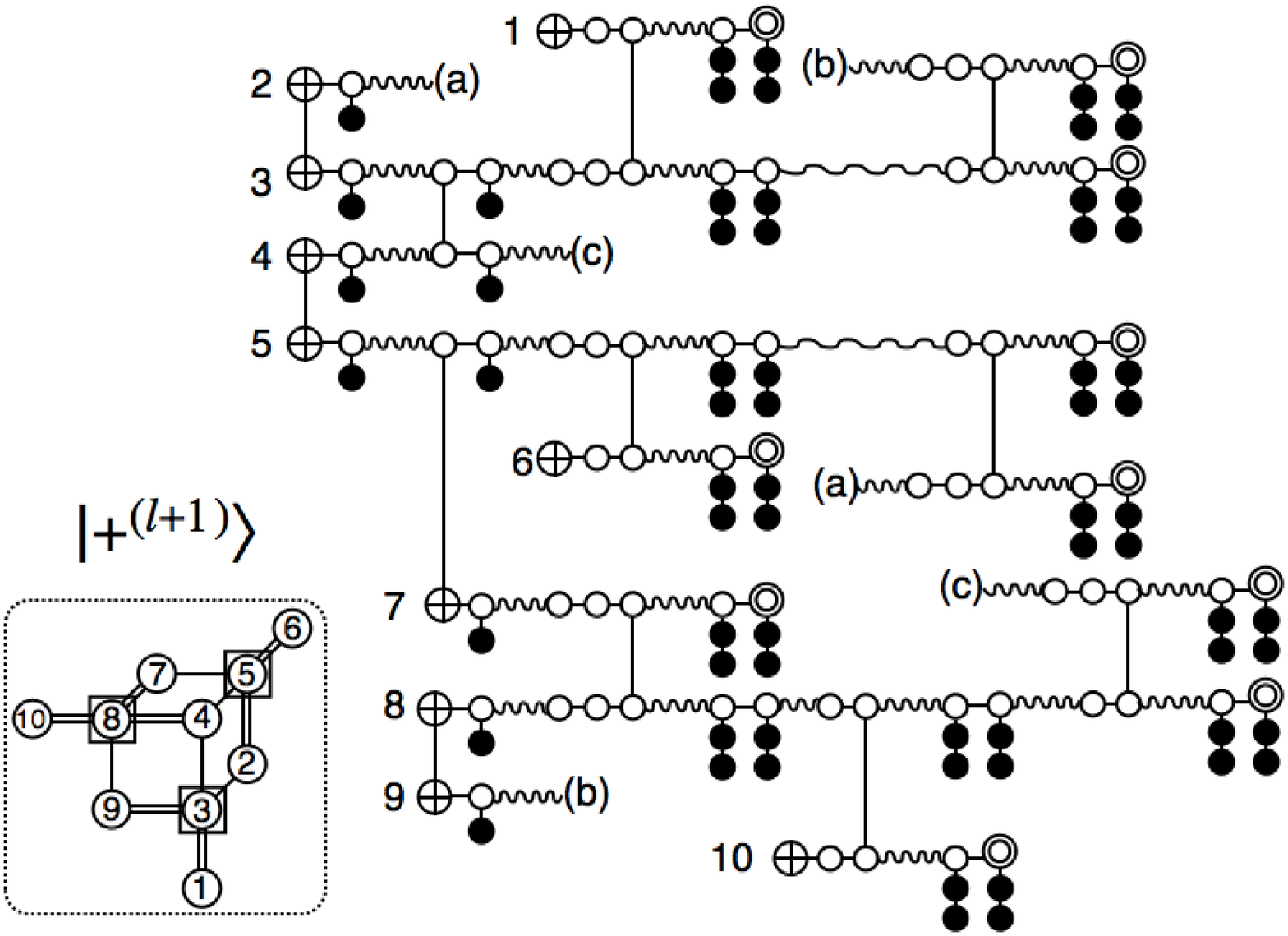}}
\label{qubit-full}
\end{equation}
\begin{equation}
\scalebox{.28}{\includegraphics*[0cm,0.5cm][29cm,22cm]{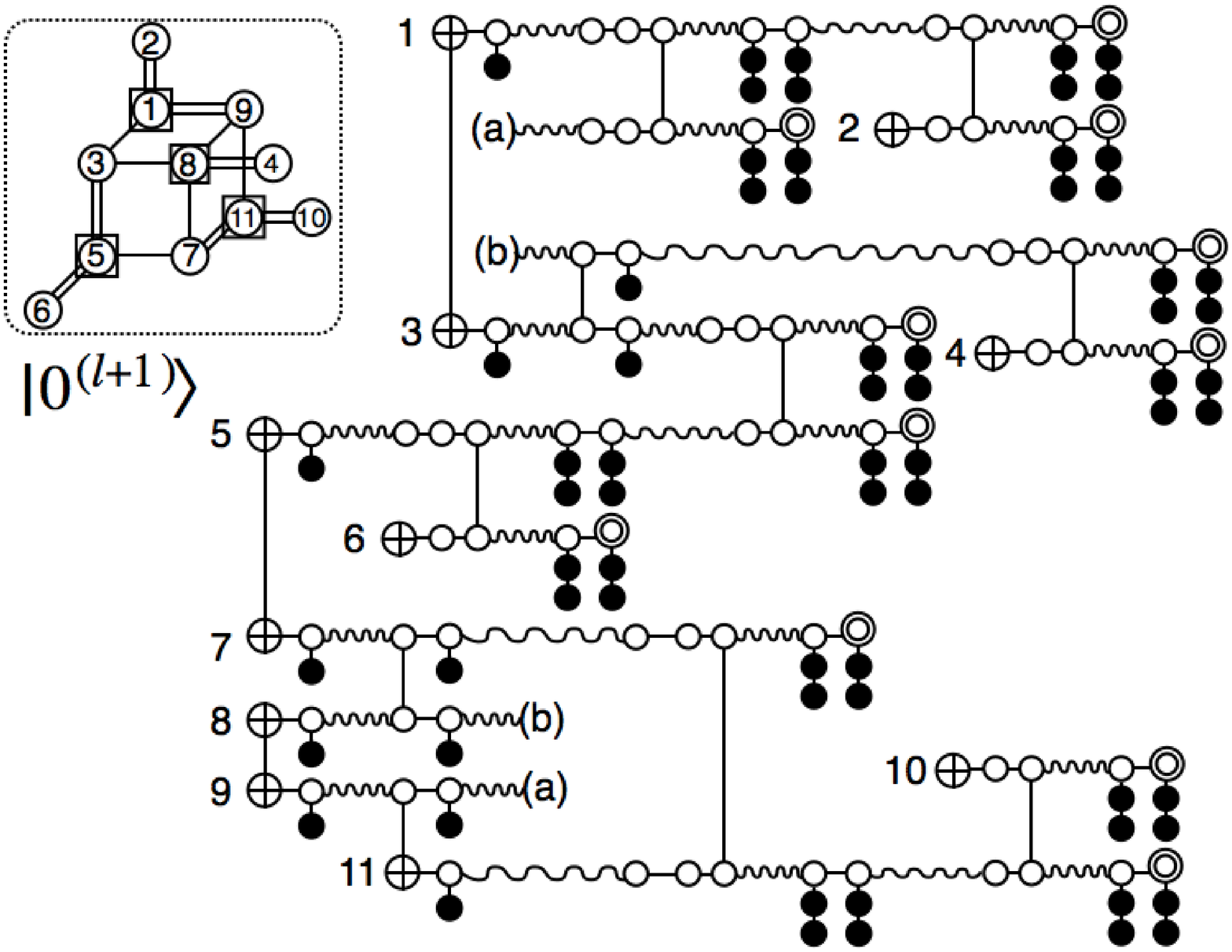}}
\label{qubit-zero-full}
\end{equation}

\end{document}